\documentclass[
reprint,
superscriptaddress,
nofootinbib,
amsmath,amssymb,
aps,
longbibliography,
]{revtex4-2}
\usepackage{graphicx}
\usepackage{dcolumn}
\usepackage{bm}
\usepackage{bbm}
\usepackage{hyperref}
\usepackage{xcolor}
\usepackage{placeins}
\usepackage[utf8]{inputenc}

\usepackage{graphicx}
\usepackage{dcolumn}
\usepackage{mathtools}
\usepackage{comment}
\usepackage{tabularx}

\usepackage{multirow}
\usepackage{enumitem}
\usepackage{array}
\usepackage{booktabs}
\newcolumntype{M}[1]{>{\centering\arraybackslash}m{#1}}
\usepackage{tikz}
\usetikzlibrary{decorations.pathreplacing,calligraphy,calc}

\newcommand{\tx}{\underline{x}}
\newcommand{\txx}{\mathbf{\underline{x}}}

\newcommand{\ty}{\underline{y}}
\newcommand{\tyy}{\mathbf{\underline{y}}}

\newcommand{\xx}{\mathbf{x}}
\newcommand{\yy}{\mathbf{y}}

\newcommand{\R}{{\mathbb R}}

\newcommand{\N}{{\mathbb N}}

\newcommand{\rhoinit}{\rho_{\textrm{init}}}
\newcommand{\rhoattr}{\rho_{\textrm{attr}}}

\newcommand{\rebel}{{conforming non-conformist }}
\newcommand{\Rebel}{{Conforming Non-Conformist }}
\newcommand{\rbl}{{CNC }}

\makeatletter 
   
\renewcommand\onecolumngrid{
\do@columngrid{one}{\@ne}%
\def\set@footnotewidth{\onecolumngrid}
\def\footnoterule{\kern-6pt\hrule width 1.5in\kern6pt}%
}

\renewcommand\twocolumngrid{
       \def\footnoterule{
       \dimen@\skip\footins\divide\dimen@\thr@@
       \kern-\dimen@\hrule width.5in\kern\dimen@}
       \do@columngrid{mlt}{\tw@}
}%

\makeatother    

\begin{document}

\preprint{APS/123-QED}
\title{Dynamical Phase Transitions in Graph Cellular Automata}

\author{Freya Behrens}
\affiliation{%
Statistical Physics Of Computation Laboratory, \'Ecole Polytechnique F\'ed\'erale de Lausanne, Lausanne, Switzerland
}%
 \author{Barbora Hudcová}%
 \affiliation{%
Algebra Department, Faculty of Mathematics and Physics,  Charles University, Prague, Czech Republic
}%
\affiliation{%
Czech Institute of Informatics, Robotics and Cybernetics, Czech Technical University, Prague, Czech Republic
}%
\author{Lenka Zdeborová}%
\affiliation{%
Statistical Physics Of Computation Laboratory, \'Ecole Polytechnique F\'ed\'erale de Lausanne, Lausanne, Switzerland
}%

\begin{abstract}
  Discrete dynamical systems can exhibit complex behaviour from the iterative application of straightforward local rules.
  A famous example are cellular automata whose global dynamics are notoriously challenging to analyze. To address this, we relax the regular connectivity grid of cellular automata to a random graph, which gives the class of graph cellular automata. Using the dynamical cavity method (DCM) and its backtracking version (BDCM), we show that this relaxation allows us to derive asymptotically exact analytical results on the global dynamics of these systems on sparse random graphs.

  Concretely, we showcase the results on a specific subclass of graph cellular automata with ``conforming non-conformist'' update rules, which exhibit dynamics akin to opinion formation.
Such rules update a node's state according to the majority within their own neighbourhood.
In cases where the majority leads only by a small margin over the minority, nodes may exhibit non-conformist behaviour.
Instead of following the majority, they either maintain their own state, switch it, or follow the minority.
For configurations with different initial biases towards one state we identify sharp dynamical phase transitions in terms of the convergence speed and attractor types.
From the perspective of opinion dynamics this answers when consensus will emerge and when two opinions coexist almost indefinitely.

\end{abstract}

\maketitle
\newpage
\onecolumngrid

\section{Introduction}

Dynamical systems can produce complex behaviour by iterating very simple local rules \cite{complexity_a_guided_tour}.
One of the simplest classes of such systems are Cellular Automata (CAs) \cite{theory_of_cellular_automata,brief_history_of_cas,wolframStatisticalMechanicsCellular1994}.
They are a popular model system due to the fascinating structures produced in the visualizations of their dynamics~\cite{wuenscheGlobalDynamicsCellular1992}.
Analysing the global dynamics of CAs is, however, notoriously difficult and many such problems are in fact proven to be undecidable \cite{undecidability_of_ca_classification, reversibility_of_2d_cas_is_undecidable}.
One aspect of the hardness comes from the fact that the regular connectivity grid of CAs imposes significant correlations between the cells.

There are numerous ways the CA regular grid structure can be relaxed to obtain a system amenable to analysis by statistical physics.
For example, the cell (or node) connectivity can be given by a random directed graph; and a (possibly different) update rule can be randomly generated for each node.
This architecture gives the synchronous, deterministic, discrete dynamical systems called Random Boolean Networks (RBNs) \cite{boolean_dynamics_with_random_couplings}.
Such a significant relaxation famously allows the RBNs' global dynamics to be analysed using mean field calculations and annealed approximations \cite{random_networks_of_automata, phase_transitions_in_random_networks, length_of_state_cycles_of_random_boolean_networks}.

In this work, we study a more subtle relaxation of the CA structure.
We consider systems where the connectivity of the nodes is determined by a random regular graph.
All nodes in this network are updated synchronously by a fixed, identical local update rule.
It is natural to call such systems Graph Cellular Automata (GCAs), although variations are known as Network Automata \cite{random_networks_of_automata}.
GCAs are very close to the CA architecture, and as such, it is still a challenge to study their dynamics analytically.
Even the annealed calculation of the number of point attractors is non-trivial compared to the RBNs due to the non-directed nature of the interactions, see e.g. \cite{dandi2023maximally}.

The main goal of this paper is to showcase a set of statistical physics tools and demonstrate that they are powerful enough to give asymptotically exact analytical results about the global dynamics of these discrete dynamical systems.
Concretely, we use the \textit{dynamical cavity method} (DCM) \cite{neriCavityApproachParallel2009,karrerMessagePassingApproach2010,mimuraParallelDynamicsDisordered2009,lokhovDynamicMessagepassingEquations2015,kanoriaMajorityDynamicsTrees2011,hwangNumberLimitCycles2020} and its \textit{backtracking} version (BDCM)~\cite{behrens2023backtracking} to give new results about the global dynamics of a specific subclass of GCAs.
This class can be intuitively understood using the terminology of opinion dynamics.
Specifically, we study GCAs with \emph{\rebel update rules}.
They have binary states~$\{0,1 \}$ and each node is updated in the following manner:
\begin{itemize}
   \item if the states in a node's neighbourhood are strongly aligned (i.e., the majority wins by at least $2\theta$ of neighbours being in the same state), the node follows the majority state in its neighbourhood
   \item otherwise, if the majority only has a slim lead over the minority (i.e., the majority wins by less than $2\theta$ neighbours being in the same state), the node gets updated in one of the following non-conformist ways:
   \begin{enumerate}
       \item[--] \textbf{type 1}  independent stubborn: the node keeps its state
       \item[--] \textbf{type 2}  independent volatile: the node changes its state
       \item[--] \textbf{type 3}  anti-conformist: the node follows the minority \label{def:type-3}
   \end{enumerate}
\end{itemize}
All nodes are updated synchronously and deterministically, using the same update rule, either of type 1, 2, or 3 for a given value of $\theta \in {\mathbb N}_0$.
The relevance of the \rebel rules stems from the fact that their dynamics can be interpreted as an opinion-formation process.

We note that the literature on opinion dynamics and its analysis through statistical physics is abundant~\cite{statistical_physics_of_social_dynamics, survey_on_nonstrategic_models_of_opinion_dynamics}.
There is a plethora of connectivity topologies and update schemes that have been studied \cite{majority_rule, axelrod_model, sznajd_model}.
Some are particularly relevant to our work, as they study the co-existence of conformist and anti-conformist behaviour \cite{homogenous_symmetrical_threshold_model_with_nonconformity, anti-conformism_in_the_threshold_model, contrarian_deterministic_effects_on_opinion_dynamics, phase_transition_in_the_majority_model}.
The type of dynamical analysis that is of relevance in the context of opinion formation dynamics is usually related to the dependence between the initial configurations and a type of attractor the system converges to.
Some exemplary questions are:
\begin{itemize}
   \item Which initial configurations can lead to consensus, and how fast?
   \item Which initial bias allows all opinions to coexist on the graph for a prolonged period of time?
\end{itemize}
\noindent
To answer these questions, we consider the \emph{density} or \textit{bias} of a binary configuration, which is its average number of 1s, and we show that various \rebel GCAs converge to qualitatively different types of attractors depending on the density of their initial configuration.

When we consider graphs with many vertices $n$, in the large system size limit, the transitions between these regions of different behaviours (e.g. finding consensus or having disagreement) become sharp: the probability to sample initial configurations that exhibit any other behaviour than what is typical for the region is going to zero.
Because the behaviours we distinguish relate to the system's dynamics, such a sharp transition is called a \textit{dynamical phase transition}.

The DCM and BDCM allow us to analytically identify values of initial densities where such a phase transition occurs. This can be confirmed by numerical experiments which show that around the phase transition, the system takes longer to converge to its typical attractor; a form of critical slowing down.
Some of the results presented here have previously been used to illustrate the BDCM in the paper that introduced the backtracking version \cite{behrens2023backtracking}.
We expand on them, by discussing their relevance in the context of cellular automata and opinion dynamics, and add results for new classes of such dynamical systems.

Concretely, we show that for multiple GCAs with \rebel rules, configurations with low initial density values almost always converge fast to the homogeneous attractor of only 0s (consensus).
However, above a certain initial density threshold, the systems instead exhibit more complex behaviour, which will be the object of our analysis with the (B)DCM.
For example, in the case of a rule always following the majority, above a certain initial density threshold the system instead converges to an attractor oscillating between two configurations of mixed states.

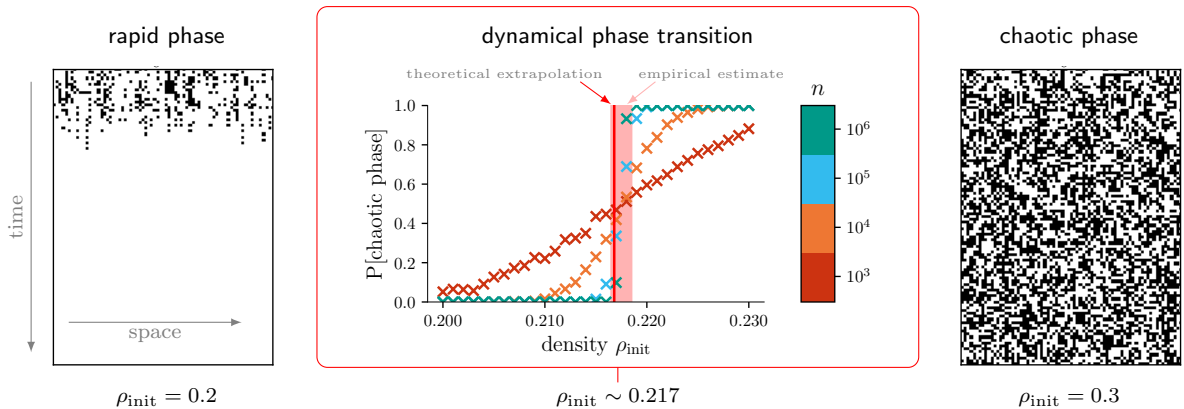
\begin{figure*}[ht]
   \centering
       \begin{tikzpicture}
       
       \node[font=\sffamily] at (-6, 2.4) {rapid phase};
       \node at (-6,0) {\includegraphics[height=4cm,trim=0.7cm 0.7cm 10.0cm 0.7cm, clip]{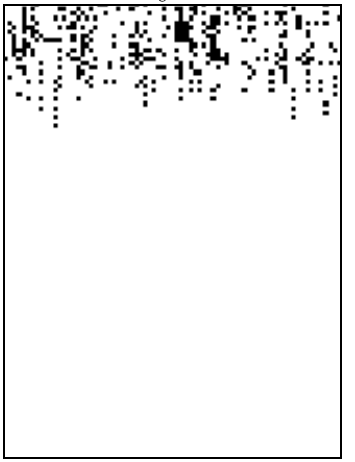}}; 
       \draw[gray, -latex] (-7.8,1.8) -- (-7.8,-1.8) node[midway, rotate=90, font=\footnotesize, yshift=6pt] {time};
       \draw[gray, -latex] (-7.3,-1.4) -- (-5.0,-1.4) node[midway, font=\footnotesize, yshift=-5pt] {space};
       \node at (-6, -2.4) {\footnotesize $\rhoinit=0.2$};

       \node at (0, 0.0) {\includegraphics[height=4cm]{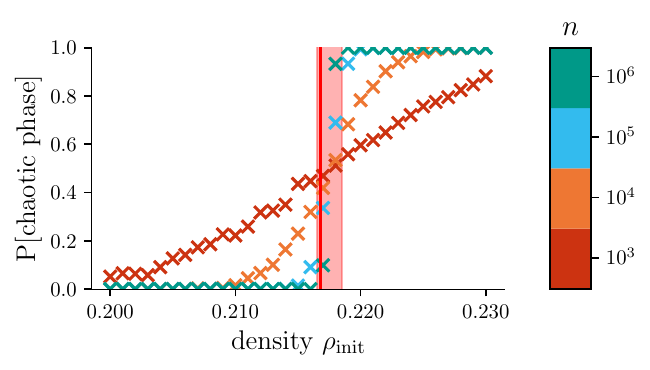}};
       \node[font=\sffamily] at (0, 2.4) {dynamical phase transition};
       \draw[red, fill=none, rounded corners] (-4,-2) rectangle (4,2.8);
       \draw[-,red] (0.0,-2) -- (0.0,-2.2);
       \node at (0, -2.4) {\footnotesize $\rhoinit\sim0.217$};
       \node[gray] at (-0.3, 1.9) {\tiny theoretical extrapolation\hspace{2em}empirical estimate};
       \draw[red, -latex] (-0.5, 1.8) -- (-0.09, 1.5);
       \draw[red!30, -latex] (0.5, 1.8) -- (0.12, 1.5);
       
       \node[font=\sffamily] at (6, 2.4) {chaotic phase};

       \node at (6, -2.4) {\footnotesize $\rhoinit=0.3$};        \node at (6,0) {\includegraphics[height=4cm,trim=8.0cm 0.7cm 2.5cm 0.7cm, clip]{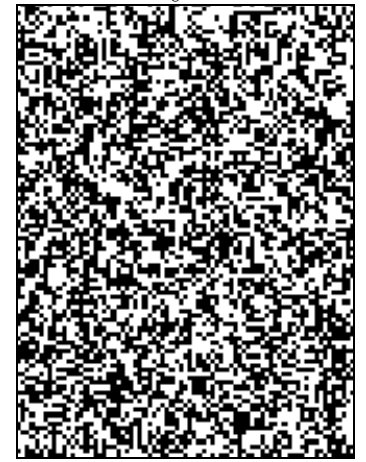}};
   \end{tikzpicture}
   
   \caption{\textbf{A phase transition diagram for a particular instance of a 5-regular GCA with a conforming anti-conformist rule $001011$.} An illustration of the system's two phases that depend on the density (i.e., the average number of black-coloured nodes) in the initial configuration.
The phases are illustrated by space-time diagrams for a system of size $n=1000$ nodes, though only a window of 75 nodes is shown.
\textit{(Left)} Rapid phase: Fast convergence to the all-0 attractor.
\textit{(Right)} Chaotic Phase: Apparent randomness in the nodes state, convergence takes longer.
\emph{(Middle)} In the large system limit, when $n\to\infty$, there is a dynamical phase transition.
At a particular initial density value $\rhoinit$, the typical behaviour of the system abruptly switches from the rapid to the chaotic phase.
For each $\rho_{\textrm{init}}$ and each system size $n$ we sampled 1024 initial configurations with the given $\rho_{\textrm{init}}$ and computed how often the system enters a chaotic phase.
For practical purposes, we conclude the system is in a chaotic phase if it does not converge within  $100*\log_2(n)$ time-steps.
The resulting frequency exhibits a sharp phase transition between $0.217$ and $0.218$, where the solid red line is our prediction from the DCM and the shaded red area comes from an empirical approximation.
This transition separates the behaviour on the left and the right.}
   \label{fig:magnum_opus}
\end{figure*}

Another interesting type of phase transition occurs for the anti-conformist rules of type 3 (following the minority instead of the majority when the race is tight).
There, as shown in Fig.~\ref{fig:magnum_opus}, for low values of initial configuration densities, the system converges to an all-0 consensus in time proportional to the logarithm of the network size.
However, above a certain initial density threshold, the system instead takes an exponentially long time to converge.

These observations bring us back to the notoriously hard-to-analyse CAs discussed at the beginning: it is a long-standing challenge in the area of discrete systems to precise the emergence of complexity \cite{complexity_measures_and_ca} and to identify a region of systems with complex behaviour \cite{edge_of_chaos}.
In multiple works on classifying dynamics of cellular automata, the typical behaviour of the system is assessed by averaging over randomly sampled initial configurations \cite{random_networks_of_automata, transition_phenomena_in_ca, measuring_complexity_using_information_fluctuation, transient_classification}.
Specific analyses with respect to the initial configuration are the exception \cite{bagnoliPhaseTransitionsCellular2014}.
Our results emphasize that for certain systems, averaging the system's behaviour over initial configurations might be a coarse process, insensitive to the particularities of different initial configuration regions.
For the anti-conformist rule we investigate, it is indeed the case that depending on the choice of initial configurations, the system either converges fast to a homogeneous attractor (simple regime), or it enters a chaotic regime.
The qualitative difference in the rule's behaviour in the two phases is significant, see Fig.~\ref{fig:magnum_opus}.

To summarize, in this work we show that the DCM and BDCM methods are powerful tools for analysing discrete dynamical systems.
We demonstrate the existence of systems with dynamical phase transitions between ordered and chaotic behaviour, and provide an analytical approach to identifying the transition between the two phases.
From the perspective of opinion dynamics, we introduce a new twist on the majority dynamics where nodes are non-conforming when the majority only has a slim lead.
Our analysis then shows how an initial bias affects the coexistence of both opinions and the time to reach a consensus or stable configuration.
From the perspective of cellular automata, we narrow the gap between the popular systems on the grid and those amenable to statistical physics.

Note that the results presented in this paper have a certain overlap with the results presented in \cite{behrens2023backtracking} by the same authors.
The paper \cite{behrens2023backtracking} was focused on the backtracking DCM that was introduced there and some of the GCAs that correspond to zero temperature dynamics in spin systems were discussed to illustrate the power of the method.
The present paper is focused on a more generic class of cellular automata and their behaviour and the BDCM together with DCM are used as methods known from the existing literature.

The paper is organized in the following manner.
Section~\ref{sec:terms} introduces all the necessary terminology regarding the dynamics of discrete systems and graph cellular automata.
Section~\ref{sec:rebel-rules-intro} introduces the \rebel rules, the dynamical phases present in such systems and showcases the dynamical phase transitions apparent from numerical experiments.
Section~\ref{sec:bdcm-method} briefly describes the dynamical cavity methods.
Section~\ref{sec:phase transitions} contains the detailed analysis of phase transitions for particular examples of the \rebel rules.

\section{Terminology and Notation}\label{sec:terms}

We call an \emph{undirected graph} of size $n$ the tuple $G = (V, E)$ where $V=\{1, \ldots, n \}$ is the set of nodes and $E=\{\{i, j \} \, | \, i, \, j \in V \}$ is the set of edges.
For each node $i \in V$ we define the \emph{neighbourhood of $i$} to be the set $\partial_i = \{j \, | \, \{i, j\} \in E \} \subseteq V$; and we define the degree of $i$ as $d(i)=|\partial_i|$.
We say an undirected graph is \emph{$d$-regular} if each node has degree $d$.
Let $G$ be a graph with $n$ nodes and let $S$ be a finite set of \emph{states}.
Each node $i$ can be assigned a state $x_i \in S$; we represent such an assignment by the sequence $\xx = x_1 \ldots x_n \in S^n$ and call it a \emph{configuration}.

\paragraph{Graph Cellular Automata.}
Let $S$ be a finite set of states.
A Graph Cellular Automaton (GCA) is a discrete dynamical system that operates on configurations of some graph with $n$ nodes.
In this work we only consider the case of random $d$-regular graphs.
The state of each node gets updated synchronously, depending on its own state and the state of its neighbours; each node uses an identical local update rule $f: S \times S^{d} \rightarrow S$.
This gives rise to a global mapping $F: S^n \rightarrow S^n$ governing the dynamics of the system.
For a configuration $\xx \in S^n$, the $i$-th node with neighbourhood $\partial_i = (i_1, \ldots, i_d)$ gets updated according to 
$$F(\xx)_i = f(x_i; x_{i_1}, \ldots, x_{i_d}).$$ 
We write a semicolon to highlight that the first entry of $f$ is always the state of the node being updated.

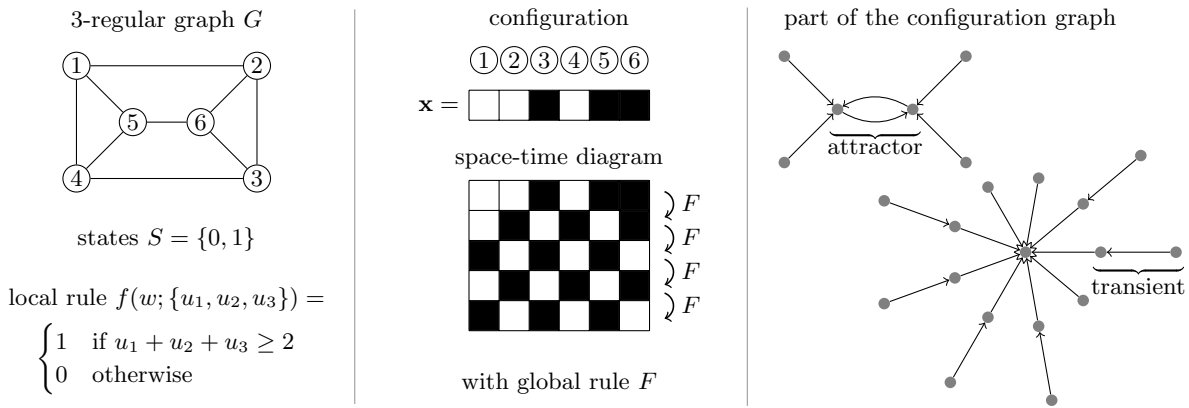
\begin{figure*}
\begin{minipage}{0.25\linewidth}
   \begin{tikzpicture}
       \node at (0,1.2) {3-regular graph $G$};
       \node[circle,draw=black,inner sep=1pt,minimum size=5pt] (1) at (-1.2,.6) {1};
       \node[circle,draw=black,inner sep=1pt,minimum size=5pt] (2) at ($(1)+(2.4,0)$) {2};
       \node[circle,draw=black,inner sep=1pt,minimum size=5pt] (3) at ($(2)+(0,-1.5)$) {3};
       \node[circle,draw=black,inner sep=1pt,minimum size=5pt] (4) at ($(3)+(-2.4,0)$) {4};
       \node[circle,draw=black,inner sep=1pt,minimum size=5pt] (5) at ($(4)+(.75,.75)$) {5};
       \node[circle,draw=black,inner sep=1pt,minimum size=5pt] (6) at ($(3)+(-.75,.75)$) {6};

       \draw [-] (1) to (2); \draw [-] (2) to (3); \draw [-] (3) to (4); \draw [-] (4) to (1);
       \draw [-] (1) to (5); \draw [-] (5) to (6); \draw [-] (6) to (2); \draw [-] (3) to (6); \draw [-] (4) to (5);
       
       \node at (0,-1.7) {states $S = \{0, 1\}$};
       \node at (0,-2.5) {local rule $f(w; \{u_1, u_2, u_3\}) =$};
       \node at (0,-3.3) {$\begin{cases}
   1 & \text{if } u_1+u_2+u_3  \geq 2\\
   0              & \text{otherwise}\end{cases}$};
\draw[-,color=black!40] (2.5,1.35) -- (2.5,-3.9);       
\end{tikzpicture}
\end{minipage}
\hspace{1.5em}
\begin{minipage}{0.28\linewidth}
   \begin{tikzpicture}
   
   \node at (1.6,1.35) {configuration};
   \node[circle,draw=black,inner sep=1pt,minimum size=5pt] (1) at (0.6,.8) {\small 1};
   \node[circle,draw=black,inner sep=1pt,minimum size=5pt] (2) at ($(1)+(.4,0)$) {\small 2};
   \node[circle,draw=black,inner sep=1pt,minimum size=5pt] (3) at ($(2)+(.4,0)$) {\small 3};
   \node[circle,draw=black,inner sep=1pt,minimum size=5pt] (4) at ($(3)+(.4,0)$) {\small 4};
   \node[circle,draw=black,inner sep=1pt,minimum size=5pt] (5) at ($(4)+(.4,0)$) {\small 5};
   \node[circle,draw=black,inner sep=1pt,minimum size=5pt] (6) at ($(5)+(.4,0)$) {\small 6};

   \def \yy{0};
   \def \y{-1.2}; 
   \def \s{0.4};

   \node at (0,\yy+0.2) {\small $\xx=$};
   \draw[step=0.4cm,color=black] (\s,\yy) grid (\s+6*\s,\yy+\s);
   \draw [fill=black] (3*\s+0.01,\yy+0.01) rectangle (4*\s-0.01,\yy+\s-0.01); \draw [fill=black] (5*\s+0.01,\yy+0.01) rectangle (6*\s-0.01,\yy+\s-0.01); \draw [fill=black] (6*\s+0.01,\yy+0.01) rectangle (7*\s-0.01,\yy+\s-0.01);

   \node at (1.6,-.5) {space-time diagram};

   \draw[step=0.4cm,color=black] (\s,\y-4*\s) grid (\s+6*\s,\y+\s);
   \draw [fill=black] (3*\s+0.01,\y+0.01) rectangle (4*\s-0.01,\y+\s-0.01); \draw [fill=black] (5*\s+0.01,\y+0.01) rectangle (6*\s-0.01,\y+\s-0.01); \draw [fill=black] (6*\s+0.01,\y+0.01) rectangle (7*\s-0.01,\y+\s-0.01);

   \draw [fill=black] (0.81,\y-\s+0.01) rectangle (1.19,\y-0.01); \draw [fill=black] (1.61,\y-\s+0.01) rectangle (1.99,\y-0.01); \draw [fill=black] (2.41,\y-\s+0.01) rectangle (2.79,\y-0.01);

   \draw [fill=black] (0.41,\y-2*\s+0.01) rectangle (0.79,\y-\s-0.01); \draw [fill=black] (1.21,\y-2*\s+0.01) rectangle (1.59,\y-\s-0.01); \draw [fill=black] (2.01,\y-2*\s+0.01) rectangle (2.39,\y-\s-0.01); 
   
   \draw [fill=black] (0.81,\y-3*\s+0.01) rectangle (1.19,\y-2*\s-0.01); \draw [fill=black] (1.61,\y-3*\s+0.01) rectangle (1.99,\y-2*\s-0.01); \draw [fill=black] (2.41,\y-3*\s+0.01) rectangle (2.79,\y-2*\s-0.01);
   
   \draw [fill=black] (0.41,\y-4*\s+0.01) rectangle (0.79,\y-3*\s-0.01); \draw [fill=black] (1.21,\y-4*\s+0.01) rectangle (1.59,\y-3*\s-0.01); \draw [fill=black] (2.01,\y-4*\s+0.01) rectangle (2.39,\y-3*\s-0.01); 
   \draw[->,semithick] (3,\y+0.25) arc[radius=.18, start angle=70, end angle=-70]; \node at (3.35, \y+0.1) {$F$};
   \draw[->,semithick] (3,\y-.2) arc[radius=.18, start angle=70, end angle=-70]; \node at (3.35, \y-.35) {$F$};
   \draw[->,semithick] (3,\y-.65) arc[radius=.18, start angle=70, end angle=-70]; \node at (3.35, \y-.8) {$F$};
   \draw[->,semithick] (3,\y-1.1) arc[radius=.18, start angle=70, end angle=-70]; \node at (3.35, \y-1.25) {$F$};
   \draw[-,color=black!40] (4.1,1.5) -- (4.1,-3.8);

   \node at (1.6,-3.5) {with global rule $F$};
   
\end{tikzpicture}
\end{minipage}
\begin{minipage}{0.3\linewidth}
   \begin{tikzpicture}
   \node at (0,1.7) {part of the configuration graph};
   \def \d{1cm}
   \def \s{0.5}
   \def \color{gray}
       \node[circle,fill=\color,scale=\s] (1) at (-1.5,.5) {};
       \node[circle,fill=\color,scale=\s] (11) at ($(1)+(135:\d)$) {};
       \node[circle,fill=\color,scale=\s] (12) at ($(1)+(225:\d)$) {};
       \node[circle,fill=\color,scale=\s] (2) at ($(1)+(\d,0)$) {};
       \node[circle,fill=\color,scale=\s] (21) at ($(2)+(45:\d)$) {};
       \node[circle,fill=\color,scale=\s] (22) at ($(2)+(-45:\d)$) {};
       
       \draw [bend right,->] (1) to (2);
       \draw [bend right,->] (2) to (1);
       \draw [->] (11) to (1);
       \draw [->] (12) to (1);
       \draw [->] (21) to (2);
       \draw [->] (22) to (2);

       \draw [decorate, decoration = {calligraphic brace}, thick] ($(2)+(.1,-.3)$) --  ($(1)+(-.1,-.3)$);
       \node at ($(1)+(\d/2, -.5)$) {attractor};

       \node[circle,fill=\color,scale=\s] (3) at (1,-1.4) {};
       \node[circle,fill=\color,scale=\s] (39) at ($(3)+(360:\d)$) {};
       \node[circle,fill=\color,scale=\s] (31) at ($(3)+(40:\d)$) {};
       \node[circle,fill=\color,scale=\s] (32) at ($(3)+(80:\d)$) {};
       \node[circle,fill=\color,scale=\s] (33) at ($(3)+(120:\d)$) {};
       \node[circle,fill=\color,scale=\s] (34) at ($(3)+(160:\d)$) {};
       \node[circle,fill=\color,scale=\s] (35) at ($(3)+(200:\d)$) {};
       \node[circle,fill=\color,scale=\s] (36) at ($(3)+(240:\d)$) {};
       \node[circle,fill=\color,scale=\s] (37) at ($(3)+(280:\d)$) {};
       \node[circle,fill=\color,scale=\s] (38) at ($(3)+(320:\d)$) {};

       \node[circle,fill=\color,scale=\s] (311) at ($(31)+(40:\d)$) {};
       \node[circle,fill=\color,scale=\s] (341) at ($(34)+(160:\d)$) {};
       \node[circle,fill=\color,scale=\s] (351) at ($(35)+(200:\d)$) {};
       \node[circle,fill=\color,scale=\s] (361) at ($(36)+(240:\d)$) {};
       \node[circle,fill=\color,scale=\s] (371) at ($(37)+(280:\d)$) {};
       \node[circle,fill=\color,scale=\s] (391) at ($(39)+(360:\d)$) {};
       
       \draw [->] (31) to (3); \draw [->] (32) to (3); \draw [->] (33) to (3); \draw [->] (34) to (3);
       \draw [->] (35) to (3); \draw [->] (36) to (3); \draw [->] (37) to (3); \draw [->] (38) to (3); \draw [->] (39) to (3);

       \draw [->] (311) to (31); \draw [->] (341) to (34); \draw [->] (351) to (35); 
       \draw [->] (361) to (36); \draw [->] (371) to (37); \draw [->] (391) to (39);

       \draw [decorate, decoration = {calligraphic brace}, thick] ($(391)+(.1,-.2)$) --  ($(39)+(-.1,-.2)$);
       \node at ($(39)+(\d/2, -.45)$) {transient};
       
\end{tikzpicture}
\end{minipage}
\caption{\textbf{Example of a GCA and its dynamics.} \emph{(Left)} Defining the GCA from a 3-regular graph $G$, state set $S$, and local rule $f$ following the majority.
\emph{(Middle)} A configuration $\xx$ (0 is white and 1 is black) and the GCA's space-time diagram starting from $\xx$.
\emph{(Right)} For the majority GCA defined on the left, we show a part of its configuration graph.}
\label{fig:terminology}
\end{figure*}

\paragraph{Global Dynamics.}
Let $F: S^n \rightarrow S^n$ be the global rule of some GCA.
We will use the symbol $\txx$ to denote a sequence of configurations from $S^n$; i.e., $\txx = (\xx^1,...,\xx^t)$ for some $t \in \N$.
If $\txx$ satisfies that $\xx^{i+1}=F(\xx^i)$ for each $i$ we call it the GCA's \emph{trajectory of length $t$ starting from the initial configuration~$\xx^1$}.
We call a matrix whose rows are configurations of a GCA at consecutive times its \emph{space-time diagram}.

Since the configuration space is finite and the update is deterministic, each long enough trajectory becomes eventually periodic.
We call the preperiod of the sequence the \emph{transient} and its periodic part the \emph{attractor} or \emph{limit cycle}.
For an attractor, the set of configurations converging to it is called its \emph{basin of attraction}.

\begin{figure*}
   \centering
   \begin{tikzpicture}
   \node at (-5,0.0) {\includegraphics[width=0.14\linewidth]{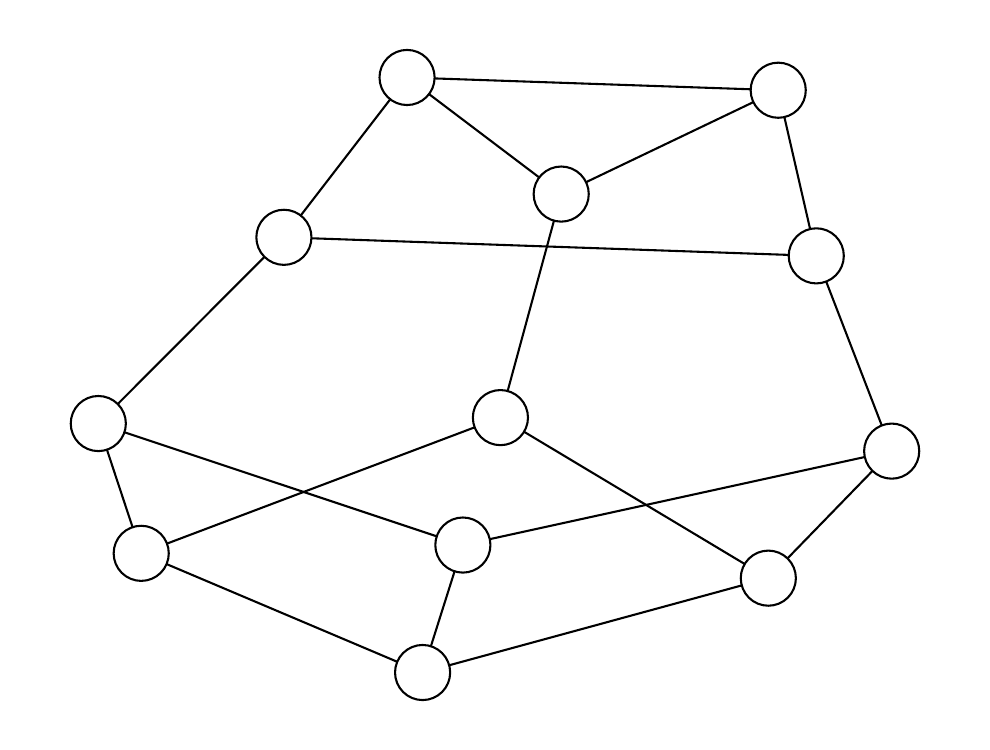} };
   \node at (-5, -1.8) {original};
   \node at (3.2, -1.8) {configuration graph};
   \draw[-,color=black!40] (-3.8,1.4) -- (-3.8,-2);
   \node at (3.2,0) {
\includegraphics[width=0.75\linewidth, trim = 2cm 1cm 2cm 1cm, clip]{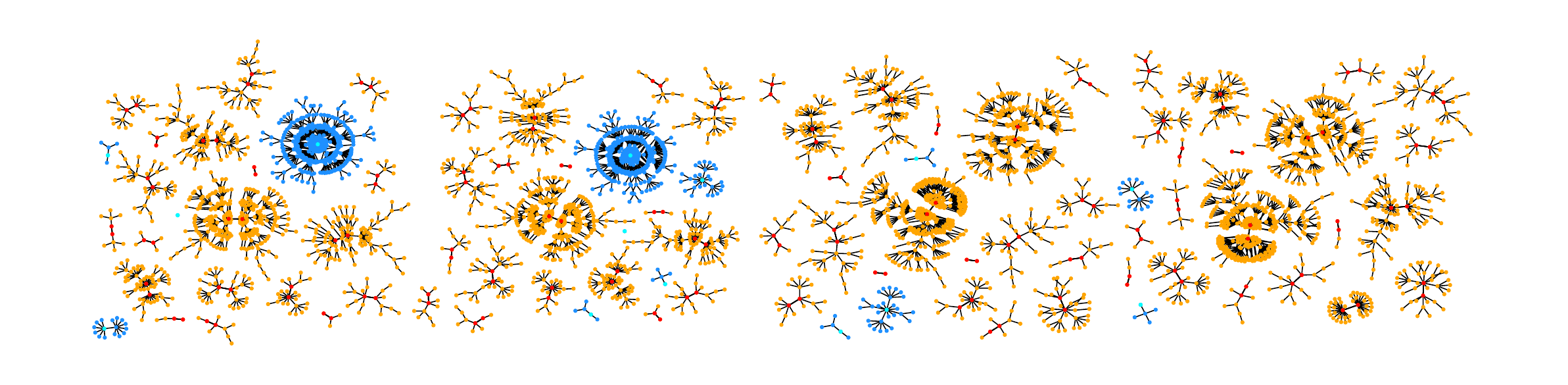} }; \end{tikzpicture}
       \caption{\textbf{Complete configuration graph of a majority GCA.} Every node is a unique configuration of the system.
The edges show how the dynamics evolve from one configuration into another.
The different colours distinguish configurations that eventually evolve into different types of attractors.
       The orange colour marks configurations leading to cyclic attractors of size 2 marked red, the blue configurations converge to point attractors in cyan.
This is the absolute majority rule, the GCA with code 0011 on a $3$-regular graph with $n=12$ nodes.
   }
       \label{fig:large_configuration_graph}
\end{figure*}
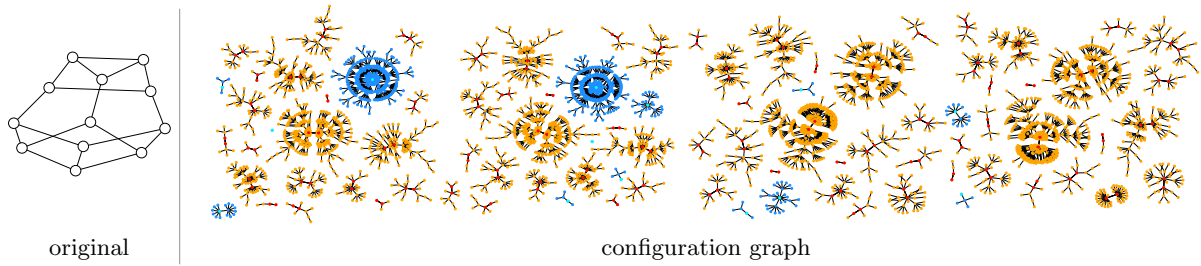

We define the \emph{configuration graph} (also called the \emph{phase space}) as an oriented graph whose vertices are the configurations from $S^n$ with edges of the form $(\xx, F(\xx)), \, \xx \in S^n$.
The notions we defined are illustrated in Figure \ref{fig:terminology} and an example of the complete configuration graph for the majority rule on a graph with 12 nodes is shown in Figure \ref{fig:large_configuration_graph}.

\paragraph{Outer Totalistic GCAs.}
A GCA is outer totalistic if its local update function ``does not distinguish between node's neighbours''.
A local rule $f$ of an outer totalistic GCA is thus a function of a node's state and the \emph{set of states} of its neighbours (oblivious to the ordering of the neighbours).
We highlight this by writing the global dynamics in the form:
$$F(\xx)_i = f(x_i; \{x_j \}_{j \in \partial_i}).$$
For example, $f: \{0,1 \}^4 \rightarrow \{0, 1 \}$ given by $f(w; u_1, u_2, u_3) = (u_1 + u_2 + u_3) \bmod 2$ gives rise to a totalistic GCA whereas copying neighbour $u_2$'s opinion given by $g(w; u_1, u_2, u_3) = u_2$ does not.

\paragraph{Outer Totalistic GCA Codes}
We restrict our study to outer totalistic GCAs with states $S = \{0,1 \}$.
In such a case, the local rule $f: \{0,1 \} \times \{0,1 \}^d \rightarrow \{0,1 \}$ is characterized by a sequence of unary Boolean functions $(f_0, f_1, \ldots, f_d)$, where for each $0 \leq k \leq d$, the function $f_k: \{0,1 \} \rightarrow \{0,1 \}$ dictates how a node changes its state if exactly $k$ of its neighbours are in state 1.
We further introduce a symbol for each unary Boolean function:
\begin{alignat*}{2} \label{eq:compressed_unary_func}
   0 &\quad \cdot \quad \text{constant 0 function}\\
   + &\quad \cdot \quad \text{identity function}\\
   1 &\quad \cdot \quad \text{constant 1 function}\\
   - &\quad \cdot \quad \text{negation}
\end{alignat*}
Thus, each local rule $f$ of an outer totalistic GCA on a $d$-regular graph is characterized by a $(d+1)$-tuple of symbols $(s_0, s_1, \ldots, s_{d}) \in \{0, 1, +, -\}^{d+1}$.
We will call this symbol sequence the \emph{code} of the rule.

For example, for $d=3$, the code $++++$ denotes the local update rule that preserves the state of each node; and the code $0011$ represents a rule that updates each node based on the majority state of its neighbours.
We note that an analogous representation has been introduced for example in \cite{outer_totalistic_cas}.

\section{\Rebel GCAs}\label{sec:rebel-rules-intro}

In this paper, we study a class of outer totalistic GCAs with \emph{\rebel update rules} (CNC).
An update rule of an outer totalistic GCA on a $d$-regular graph with states $S=\{0,1\}$ is \rbl{}with threshold $\theta \in \N_0$ if it updates each node in one of the following ways:
\begin{itemize}
   \item \textbf{strong agreement region: } if the majority wins by at least $2\theta$ of nodes; i.e., $|\sum_{j \in \partial i} x_j - \frac{d}{2}| \geq \theta$; the node conforms to the majority of its neighbours
   \item \textbf{weak agreement region: } if the majority wins by less than $2\theta$ of nodes; i.e., if $|\sum_{j \in \partial i} x_j  - \frac{d}{2}| < \theta$; the node gets updated in a non-conformist way:
      \begin{itemize}
          \item stubborn independent: the node keeps its state; code type ``\textbf{0+1}''
          \item volatile independent: the node changes its state; code type ``\textbf{0$-$1}''
          \item anti-conformist: the node follows the minority of its neighbours; code type ``\textbf{0101}''
       \end{itemize}
\end{itemize}
All the nodes in the network get updated synchronously, using the same update rule, either of type \textbf{0+1}, \textbf{0$-$1}, or \textbf{0101}. As an example, for $d=5$, the anti-conformist GCA with threshold $\theta=1$ corresponds to the rule with code $001011$, and $\theta=2$ gives the rule with code $011001$.

We note that an odd connectivity $d$ and $\theta=0$ imply that all neighbourhood configurations result in a strong agreement region.
In such a case, a node always conforms to the majority 
and this gives the well-studied case of absolute majority rules with code type ``\textbf{01}''.
Whenever $\theta \geq 1$, some neighbourhood configurations result in a weak agreement region where the rules \textbf{0+1}, \textbf{0$-$1}, or \textbf{0101} demonstrate different forms of non-conformist behaviour.

The anti-conformist case of \rbl{}rules has a particularly interesting interpretation in the context of opinion making: if the agreement of one's neighbours is weak, one has enough ``courage'' to demonstrate an attitude different from the majority. However, once the neighbours' opinion alignment is too strong, one conforms to the opinion of the majority.

\paragraph{Short Attractors.}
An important property of the \rbl{}GCAs is that for an arbitrary system size, they only seem to have short attractors.
As we will see, this is a crucial property that allows us to apply the BDCM method and analyse properties of the most typical attractor of the \rbl{}GCAs.  

Specifically, the absolute majority rules, together with the stubborn and volatile independent rules belong to a wider class of \emph{majority threshold rules} which, irrespective of the system size, only have attractors of size 1 and 2.
This applies to an arbitrary topology of the connectivity network, as long as it has undirected edges.
This has been proved in \cite{decreasing-energy_functions} using an elegant argument by introducing a decreasing energy function for such systems.

For the case of anti-conformist \rbl{}rules, we so far lack a proof of such a property.
However, the numerical results suggest that attractors larger than 2 are not typical for anti-conformist GCAs of large size, as we only rarely sampled them (Appendix, Fig.~\ref{app:fig:cycle-lengths}).
We note that the topology of a random regular graph seems crucial here as for preliminary experiments on a regular grid we encountered attractors larger than~2.

\paragraph{Related Work.}
The class of \rbl{}rules, seemingly simple, contains systems with a wide variety of behaviour that have received a lot of attention in the literature, although not always in exactly the synchronous setting on random regular graphs.
The interest is due to the rules' relevance in different application fields.
For cellular automata, typically on lattices, density classification is used as a vehicle for reasoning about their computational capabilities \cite{schonmannBehaviorCellularAutomata1992,busicDensityClassificationInfinite2012}.
Bootstrap-percolation \cite{phase_transitions_in_twoway_bootstrap_percolation,altarelliLargeDeviationsCascade2013} or the zero-temperature Glauber dynamics \cite{morrisZerotemperatureGlauberDynamics2011,damronZerotemperatureGlauberDynamics2019} can also be modelled with \rbl{}rules and are studied on various types of graphs.

\begin{figure*}
   \centering
   \begin{minipage}{0.9\linewidth}
   \begin{tikzpicture}
       \node at (8,0) {\includegraphics[width=0.43\linewidth]{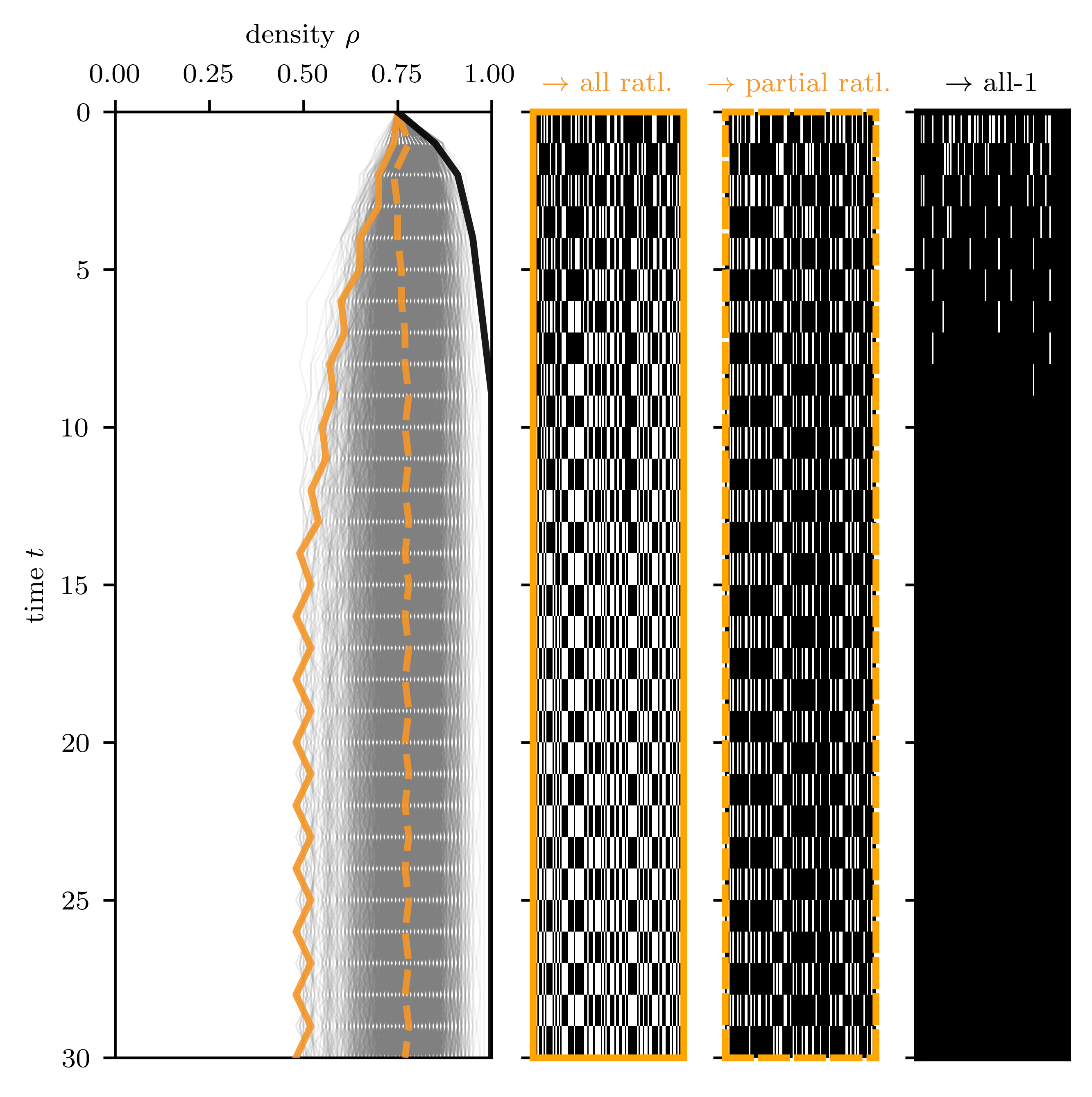}};
       \node at (0,0) {\includegraphics[width=0.43\linewidth]{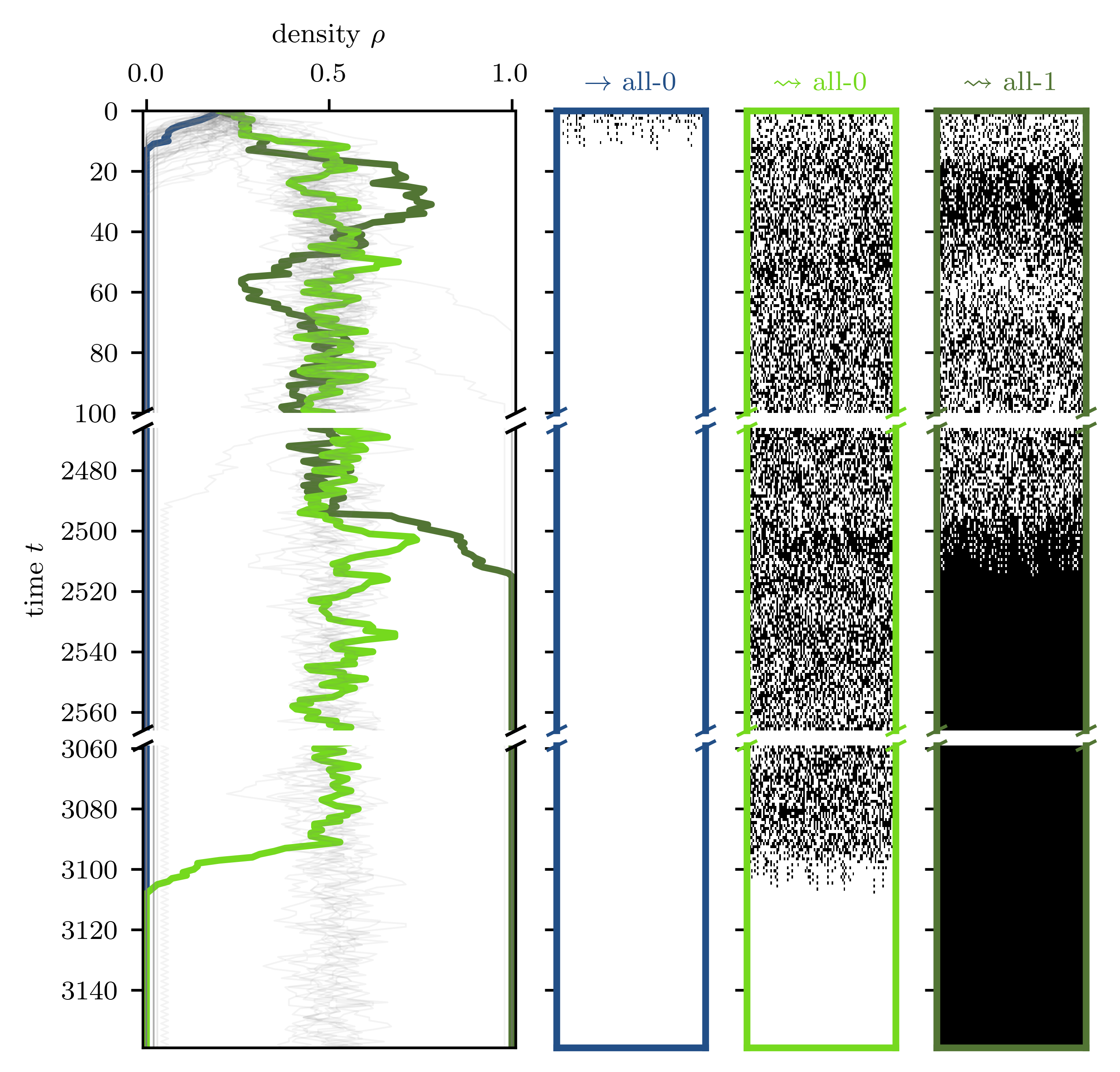}};
       \node[font=\sffamily] at (0,4) {{anti-conformist: $001011$}};
       \draw[-] (4,-4) -- (4,4.5);
       \node[font=\sffamily] at (8,4) {{volatile independent: $00$$-$$11$}};
   \end{tikzpicture}
   \end{minipage}
   \caption{\textbf{Two examples for CNC dynamics.} We show two examples for \rebel dynamics on small random regular graphs with $n=100$ nodes. We sample initial configurations for a fixed initial density $\rho$ and show how they evolve (light gray lines). Some samples are highlighted in color and their corresponding space-time diagrams are shown. \textit{(Left)} The anti-conformist GCA $001011$ is started from $\rho=0.2$, in this case, the time axis is broken for visualization purposes, as for some samples the time to an attractor is extremely long. We highlight three different behaviours: Two samples converge the the configuration with only 0s (blue), but one takes a very long time to reach it (light green) and one converges rapidly to a configuration with only 1s (dark green).
\textit{(Right)} The volatile independent GCA $00$$-$$11$ if started from $\rho=0.75$. Here the orange solid line represents the dynamics of a sample where eventually all nodes change their state in an length-2 limit cycle. We call such nodes rattling (ratl.) The dashed line shows a sample which is partial rattling, i.e. some nodes are stable in the attractor, but some are rattling. Finally, the black line is a sample which ends up in the attractor with only 1s.}
   \label{fig:example-rules}
\end{figure*}
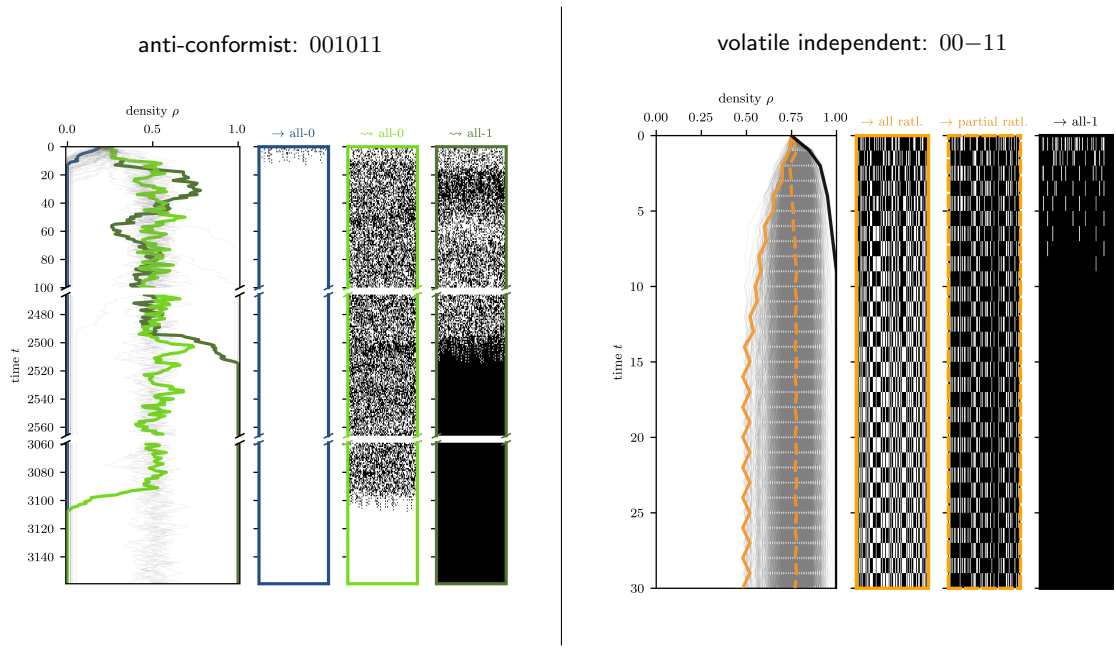

The \rbl{}rules also play a prominent role in modelling opinion spreading.
The co-existence of conformist and anti-conformist dynamics has been studied in models of collective behaviour \cite{anti-conformism_in_the_threshold_model, homogenous_symmetrical_threshold_model_with_nonconformity}.
However, the co-existence is typically introduced in one of the two following ways:
\begin{enumerate}
   \item The network consists of two types of nodes, conformist ones that always follow the majority and anti-conformist ones always following the minority.
   \item With probability $p$ a node gets updated using a majority rule, and with probability $1-p$ it gets updated in an anti-conformist way.
\end{enumerate}
In contrast, for the \rebel rules as considered in the present paper, the behaviour of a node is entirely determined by the nodes in its neighbourhood, not by external probabilities.
We show two examples of such dynamic behaviour in Figure~\ref{fig:example-rules}, where for the anti-conformist GCA $001011$ and the volatile independent GCA $00$$-$$11$ we show three different initializations and their long-time behaviour in space-time diagrams.

The connection between the CAs and opinion dynamics on graphs is discussed in \cite{bagnoliPhaseTransitionsCellular2014}.
All the mentioned applications directly raise relevant questions on the dynamics, e.g. how quickly or if at all one can reach consensus given an initial configuration \cite{pelegLocalMajoritiesCoalitions2002,kanoriaMajorityDynamicsTrees2011}.
In the following, we show how to answer such questions for these seemingly simple but ubiquitous rules.

\subsection{Types of Dynamical Phases}
For \rebel GCAs we identify a number of qualitatively different phases the system exhibits when varying the density of 1s in the initial configuration.
For the transients, we distinguish phases of slow and fast convergence.
For the attractors, we distinguish between attractors of size 1 and 2, between the density of 1's in the attractor's configurations and the portion of nodes that are changing their state in a cyclic attractor.
We call a specific combination of a transient and attractor type a \textit{dynamical phase}.
A \textit{dynamical phase transition} is an abrupt, non-analytic change from one dynamical phase to another.
It is the critical point where the system exhibits different qualitative behaviours on either side of the transition.
This is defined in the large $n$ limit, when the system has many interacting nodes.

To define this formally, let $\txx=(\xx^1,...,\xx^p,...,\xx^{p+c})$ be a trajectory of a threshold GCA with a transient of length $p$ leading into an attractor of length $c$.

\paragraph{Initial configuration.} We define the \emph{density} or \textit{bias} of a configuration $\xx \in \{0,1 \}^n$, $\xx = (x_1, \ldots, x_n)$, as:
\begin{align}
   \rho(\xx) = \frac{1}{n} \sum_{i=1}^n x_i.
\end{align}

The \textit{initial density} for the trajectory $\txx$ is $\rhoinit(\txx) = \rho(\xx^1)$.
We will show that as we vary $\rhoinit$, the system exhibits changes in the phase it converges to that become more and more abrupt as the system size grows $n\to \infty$.

\paragraph{Transient types.} We say that the convergence to an attractor is rapid (ordered), if the transient length $p$ as a function of the system size $n$ grows as $O(\log n)$. Similarly, convergence is chaotic if it takes a long time, namely $p$ grows in $\Theta(\exp n)$. We conjecture from the numerical investigations that intermediate transient lengths do not appear in the systems considered here.

\paragraph{Attractor types.}
In general, we define the density of a limit cycle/attractor of length $c$ as the average density over all its configurations:
\begin{align}
   \rho_{\textrm{attr}}(\txx) = \frac{1}{c} \sum_{t=1}^c \rho(\xx^{p+t}).
\end{align}
For all attractors of size $c > 1$, we say that the $i$-th node is a \emph{rattler} if it changes its state at least once in the limit cycle. Otherwise, we say that the $i$-th node is stable. We define the \textit{activity} of a limit cycle as the average number of its rattlers, formally:
\begin{align}
\alpha(\txx) =&\frac{1}{n}\sum_{i\in V} \mathbbm{1}\left[1 \leq \sum_{t=p+1}^{p+c-1}\mathbbm{1}[ x^t_i \neq x^{t+1}_i]\right]\label{eq:activity}
\end{align}
With these definitions, we distinguish the four attractor types in Table~\ref{tab:attractor-types}.
\begin{table*}
   \small
   \centering
   \begin{tabular}{ M{.8cm} m{2.75cm} m{8cm} m{2cm} } 
    \toprule
icon & attractor & description & parameters \\ [0.5ex] 
\toprule
\vspace{0.1cm} \includegraphics[width=0.02\textwidth]{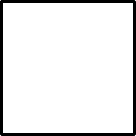} & \multirow{2}{2.75 cm}{\textit{homogeneous stable}} & \multirow{2}{8 cm}{almost only point attractors with almost all nodes in state 0 or almost all nodes in state 1} & \multirow{2}{2 cm}{$c=1$\\$\rho_{\mathrm{attr}} \in \{0,1 \}$} \\ 
 \includegraphics[width=0.02\textwidth]{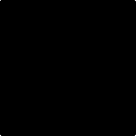} & &  &  \\
\hline
\vspace{0.1cm} 
\multirow{2}{*}{\includegraphics[width=0.02\textwidth]{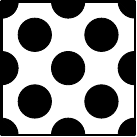}} & \multirow{2}{2.75 cm}{\textit{mixed-colour stable}} & \multirow{2}{8 cm}{almost only point attractors where at least a constant fraction of both 0's and 1's is present } & \multirow{2}{2 cm}{$c=1$\\$\rho_{\mathrm{attr}} \in (0,1 )$} \\ 
& & &  \\
\hline
\vspace{0.1cm} 
\multirow{2}{*}{\includegraphics[width=0.02\textwidth]{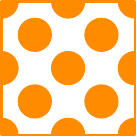}} &  \multirow{2}{2.5 cm}{\textit{partially rattling}} & \multirow{2}{8 cm}{almost only 2-cycles with at least a constant fraction of both rattling and stable nodes} & \multirow{2}{2 cm}{$c=2$\\$\alpha \in (0,1)$} \\ 
& &  &\\ 
\hline
\vspace{0.1cm} 
\multirow{2}{*}{\includegraphics[width=0.02\textwidth]{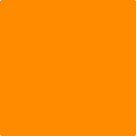}} & \multirow{2}{2.5 cm}{\textit{all-rattling}} & \multirow{2}{8 cm}{almost only 2-cycles with almost all rattling nodes}   & \multirow{2}{2 cm}{$c=2$\\$\alpha = 1$} \\ 
& &  &   \\
\hline
\end{tabular}
   \caption{Four types of attractors, marking different destinations of their dynamical behaviour. We emphasize that our definition makes the distinction for $\alpha$ and $\rho_{\mathrm{attr}}$ only up to to a finite fraction $\Theta(n)$
   of the nodes. This disregards a subleading number $o(n)$
   of nodes that might have a different state in the homogeneous stable attractor, or $o(n)$ nodes that are not rattling in the all-rattling attractor. Likewise, the phases ignore $o(n)$ of transients which converge to attractors with limit cycle lengths with $c \notin \{1,2\}$. (Informally, $g(n) \in \Theta(f(n))$ if $g$ grows with the same order as $f$ and  $g(n) \in o(f(n))$ if $g$ grows slower than $f$.)} 
   \label{tab:attractor-types}
\end{table*}

\begin{figure*}[htbp!]
   \centering
   \includegraphics[width=1\linewidth]{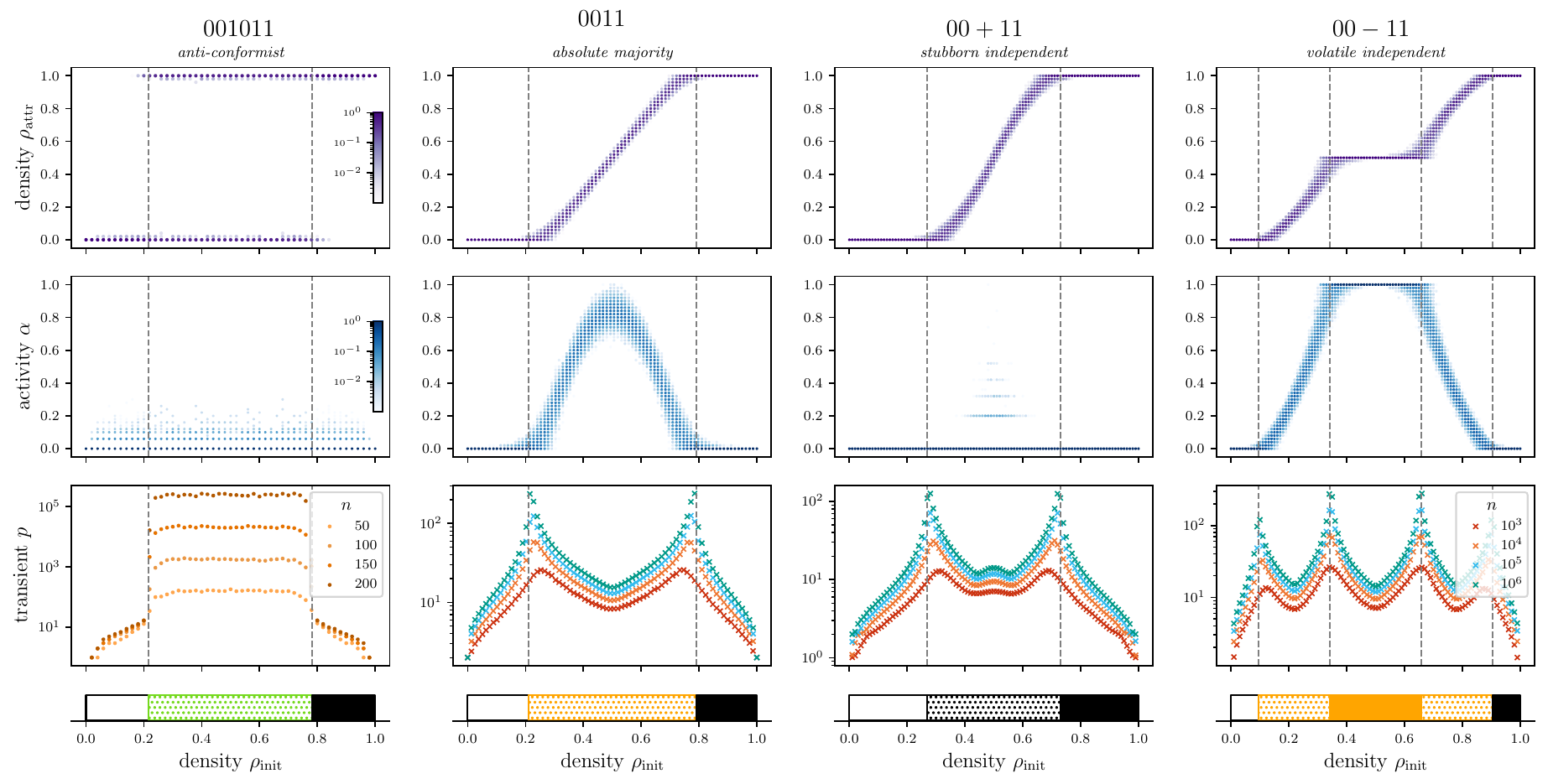}
   \caption{\textbf{Numerical experiments for four types of CNC rules for $d=3,4,5$.} For rules $0011$, $00$$+$$11$ and $00$$-$$11$ we sampled $1024$ graphs for every $n \in \{ 10^3,10^4,10^5,10^6\}$ and every initial density $\rho_{\mathrm{init}} = \frac{k}{100}, \, k \in \{0, 1, \ldots, 100\}$.
   For $001011$, due to the exponential explosion of the transient length in $\sim (0.2,0.8)$, we used $n \in \{50,100,150,200\}$.
   \emph{(First and second row)} Histograms of the properties of attractors: their density and the fraction of rattlers.
   With the exception of GCA $001011$, they were computed for $n=10^3$ with a binning on the y-axis for both
   $\rho_{\mathrm{attr}}$
   and $\alpha$ with $101$ bins. For the GCA $001011$ graphs of size $n=200$ and $51$ bins on the y-axis were used.
   \emph{(Third row)} Average transient length $p$ for $0011$, $00$$+$$11$ and $00$$-$$11$; median transient length for $001011$. We observe behaviour consistent with either exponential or logarithmic growth of the transient lengths as a function of the system size.
   \emph{(Last row)} Diagram showing the dynamical phases corresponding to each of the attractor and transient type. Transitions between the phases correspond to peaks in the transient length or a change in the scaling of the transient regime. We use the  color schemes from Table~\ref{tab:attractor-types} when the convergence to the attractor is rapid. The light green color denotes a slow convergence. Note that the same data for the rules $00$$+$$11$ and $00$$-$$11$ were already used in \cite{behrens2023backtracking} to illustrate the results that the BDCM can obtain.}
   \label{fig:overview-empirics}
\end{figure*}

\paragraph{Empirical Locations of Dynamical Phases.}
On finite systems, we can empirically measure all the previously defined properties and their scaling in the graph size $n$.
For now, we explore GCAs with four rules: The absolute majority rule, and three rules with the different possible non-conforming behaviours under weak agreement (stubborn, volatile independent, and anti-conformist).
Fig.~\ref{fig:overview-empirics} shows the transient length scaling in $n$, the attractor's density $\rho_{\textrm{attr}}$ and activity $\alpha$ in terms of initial density $\rho_{\textrm{init}}$.

Clearly, for all four rules the homogeneous all-zero and all-one state is an attractor of the dynamics.
We observe that when the initial bias is close to such a homogeneous attractor the convergence to it is rapid for all four rules.
All the studied rules undergo a phase transition for some value $\rhoinit$ which jointly occurs with a slowing down of the convergence (an increase in transient lengths).

The anti-conformist rule's behaviour stands out, where the exponentially long transients lead to the all zero or all one attractor with equal probability.
For the other three rules (majority, stubborn/volatile independence) the slowing down is within the $O(\log n)$ regime, just with a larger growing prefactor in the vicinity of the phase transition. For the volatile independence rule, there are even two such transitions.
The transient behaviour for anti-conformist rule is very different.
It switches from the short $O(\log n)$ to the long $\Theta(\exp n)$ transients around the critical point close to $\approx 0.2$ and $\approx 0.8$. The long transients are maintained throughout the dynamical phase.

The different phases confined by those transitions occur as follows:
The absolute majority rule on the $3$-regular graphs phase rapidly converges to the partially rattling state, where a core is stable, and some nodes are changing their opinion at every iteration.
The stubborn independent rule on the $4$-regular graph produces an attractor with mixed (i.e. 0 and 1) stable opinions which is reached rapidly.
The volatile rule, coming from the homogeneous all-zero attractor and increasing $\rhoinit$, first goes through a dynamical phase of rapid convergence towards a partially rattling state, similar to the majority rule before. For very weak initial bias, eventually all nodes keep switching their colors -- the all-rattling attractor.

Previous work identified similar dynamical phases for the threshold $q$-voter model \cite{vieiraThresholdVoterModel2018}, although that work considers a thresholded, noisy version of the majority rule.

We further highlight that the transitions only become sharp for large $n$. For smaller finite systems and particular initial density values $\rhoinit$,  we can observe the co-existence of phases at both sides of the transitions. For example, this happens for the two GCAs on small graphs with $n=100$ nodes that are shown in Fig.~\ref{fig:example-rules}, Section~\ref{sec:rebel-rules-intro}.

In Appendix~\ref{app:sec:large-d} we provide some empirical results for examples with larger degrees.
For rules which belong either to the absolute majority, stubborn and volatile independent rules, scaling the threshold $\theta$ as $O(1/\sqrt{d})$ exhibits the same transitions as the degree $d$ grows, consistent with the type of large $d$ behaviour observed in \cite{behrensDisAssortativePartitions2022a}.
For the anti-conformist the picture is less clear as new types of behaviour emerge that are different from what we observed for the GCA $001011$.
Overall, we leave thorough empirical and theoretical investigations of larger degrees and their appropriate parameterizations to future work.

In the remaining Sections, we supplement our empirical results with a theoretical analysis of the precise positions of the phase transitions. For this, we first introduce the (backtracking) dynamical cavity method in Section~\ref{sec:bdcm-method}, and then present the derived analytical dynamical phase transitions in Section~\ref{sec:phase transitions}.

\section{Dynamical Cavity Methods}\label{sec:bdcm-method}

To analyse the dynamics of the previously introduced family of CNC GCAs, we use the \textit{dynamical cavity method} (DCM) \cite{neriCavityApproachParallel2009,karrerMessagePassingApproach2010,mimuraParallelDynamicsDisordered2009,lokhovDynamicMessagepassingEquations2015,kanoriaMajorityDynamicsTrees2011,hwangNumberLimitCycles2020} and its extension, the \textit{backtracking dynamical cavity method} (BDCM) \cite{behrens2023backtracking}.
These methods are inspired by the cavity method from statistical physics which has proven its success in the analysis of static systems \cite{mezard2003cavity}.
While their results hold in the thermodynamic limit, i.e. when the number of nodes $n$ tends to infinity, we will see that the behaviour of systems with relatively small $n$ already corresponds well to the theoretical predictions for large $n$.

Both methods consider motifs from the configuration graph (Fig.~\ref{fig:large_configuration_graph}) that represent dynamical phenomena as the static element of a cavity analysis.
The idea of the DCM is to take finite trajectories from the configuration graph.
Similarly, the BDCM considers finite trajectories that lead into cycles of a fixed length.
A general motif that encompasses both ideas is the \textit{$(p/c)$ backtracking attractor}, defined as \begin{align}\txx=(\xx^1, \xx^2, \ldots, \xx^p, \xx^{p+1}, \ldots, \xx^{p+c}) \in (S^n)^{p+c}\,,\end{align} for $p, c \in \mathbbm{N}$ where the first $p$ configurations compose a transient and the last $c$ configurations a limit cycle.
Therefore, $c=0$ gives the trajectories without attractors for the DCM and $c > 0$ gives the BDCM.
Despite the static methodology, the backtracking attractor is inherently dynamic, so the static analysis allows one to infer back results about the dynamics.
In order to identify the dynamical phase transitions from the previous section, it suffices to answer the following question: What are the average properties of the typical (= most numerous) backtracking attractor for a fixed $\rhoinit$ when $p\to\infty$?

\paragraph{Introduction to (B)DCM.}
Before we answer this question precisely for the \rebel GCAs we give a brief overview to the (B)DCM, to make clear how it works - and why this approximation is valid for the \rebel rules on random regular graphs.
In this introduction, we want to give a good understanding of the method. However, we refer the reader to \cite{behrens2023backtracking} for the original derivation.

The main ingredient to the (B)DCM is a probability distribution over all possible sequences of configurations $(S^{n})^{p+c}$.
In the simple case, the probability assigns a uniform value to all $(p/c)$ backtracking attractors $\txx$ that occur in the configuration graph of the dynamics, and a zero measure to any other sequence:
\begin{eqnarray}
   P(\txx) \!= \!\frac{1}{Z}  \mathbbm{1}\left[F(\xx^{p+c}) = \xx^{p+1}\right]\! \prod_{t=1}^{p+c-1}  \mathbbm{1}\left[F(\xx^{t}) = \xx^{t+1}\right]. \label{eq:prob_dist}
\end{eqnarray}
Here, $\mathbbm{1}(\cdot)$ is the indicator function on a Boolean statement where a true statement yields 1 and 0 otherwise.
If $c=0$ and therefore $x^{p+1}$ is undefined, we drop the first factor where it appears.
The normalization constant $Z$ of this distribution is then equivalent to the number of valid backtracking attractors.
Since this number $Z$ is extensive in the system size $n$, we measure it in terms of the free entropy density $\Phi = \frac{1}{n}\log(Z)$.
However, computing $Z$ and therefore the entropy directly is intractable, due to the high-dimensional integral over all possible configurations.
To solve this issue, analogous to the classical cavity method for static analysis, we use the Bethe Peierls approximation to compute its leading exponential factor using Belief Propagation (BP) on its factor graph. This approach is exact for factor graphs that are trees and, in many cases, leads to asymptotically exact results for sparse locally tree-like factor graphs. In the literature, the cases where the BP provides asymptotically exact results on sparse random graphs are called replica symmetric and \cite{behrens2023backtracking} observed that it indeed plausibly provides asymptotically exact results for the cases studied there.

Eventually, this approach leads to a lower dimensional fixed point equation which is amenable to numerical solutions.
In addition to the approximation of the free entropy density $\Phi_{BP}$, this approximation conveniently admits a means of computing the marginals of the probability distribution in \eqref{eq:prob_dist} and expectations for observables\footnote{This is only possible when the observable factorizes over the nodes.} of the system, e.g. the density of the attractor.
By additionally introducing re-weighting of the backtracking attractors in the probability distribution according to some external potential we can also `fix' some of their properties to a prescribed constraint, and extract for example only backtracking attractors with a fixed initial density $\rhoinit$.

\paragraph{Equations for random regular graphs.}
For random $d$-regular graphs this strategy admits a particularly simple analysis: Under the assumption that all neighbourhoods are locally the same, solving the BP on the factor graph corresponding to eq.~\eqref{eq:BP-equation} is equivalent to solving a fixed point equation for only one neighbourhood.
Then, the message on the factor graph $\chi_{\tx,\ty}^{\to} \in \R^{4(p+c)}$ from the center node $x$ to its neighbour $y$ is defined in terms of all possible values that its other $d-1$ neighbours $\yy$ can take.
It is re-weighted by $\chi^{\to}$ itself~\footnote{This equation is equivalent to (17) from \cite{behrens2023backtracking}, and the derivation and factor graph is described therein using the same notation.}:
\begin{align}
   \chi_{\tx,\ty}^{\to} = \frac{1}{Z^{\to}} \underbrace{\vphantom{\prod_{t=1}^{p+c-1} } e^{-\lambda \tilde{\Xi}(\tx)}}_{\substack{a(\tx)\\\mathrm{observable/}\\\mathrm{constraint}}} \sum_{\tx, \tyy_{[d-1]}}\left( \underbrace{ \mathbbm{1}\left[  f(x^{p+c};\yy^{p+c}_{[d]}) = x^{p+1}\right] \prod_{t=1}^{p+c-1}  \mathbbm{1}\left[  f(x^t;\yy^t_{[d]}) = x^{t+1}\right]}_{\substack{\mathcal{A}(\tx,\tyy_{[d]}) \\\textrm{valid $(p/c)$-backtracking attractor}\\ }} \prod_{\underline{z} \in \tyy_{[d-1]}} \chi_{\underline{z}, \tx}^{\to}\right).\label{eq:BP-equation}
\end{align}
Here, the inner constraint assures that we only consider valid backtracking attractors. The $Z^{\to}$ is again the normalization constant, the interval is $[k]=1,...,k$ and $\tilde{\Xi}$ is the factorized observable of the global extensive variable of interest $\Xi(\txx) =\frac{1}{n}\sum_{i=1}^n \tilde{\Xi}(\tx)$.
This localized observable $\tilde{\Xi}$ with the factor $\lambda$ allows for the previously mentioned re-weighting and constraining. As an example, take the initial density, for which we define the terms of the summand $\tilde{\Xi}(\tx)=x^1$, so that the intensive global variant is $ \frac{1}{n}\Xi(\txx) =\frac{1}{n}\sum_{i=1}^n \tilde{\Xi}(\tx) = \rhoinit(\txx)$.

To obtain the BP approximation of the entropy density it suffices to compute the following at the fixed point of \eqref{eq:BP-equation}:
\begin{align}
   \Phi_{\mathrm{BP}} &= \log(Z^{\mathrm{fac}}) - \frac{d}{2} \log(Z^{\mathrm{var}})\,,\label{eq:BP-entropy}\\
   Z^{\mathrm{fac}} &= \sum_{\tx, \tyy_{[d]}} \mathcal{A}(\tx,\tyy_{[d]}) \prod_{\ty \in \tyy_{[d]}} \chi_{\tx, \ty}^{\to}\,,\\
   Z^{\mathrm{var}} &= \sum_{\tx, \ty} a(\tx) \chi_{\ty,\tx}^{\to} \chi_{\tx,\ty}^{\to} \,.
\end{align}
Implementing and finding a solution to \eqref{eq:BP-equation} can be non-trivial due to numerical instabilities.
The solver used for our analysis is available on github\footnote{\url{github.com/SPOC-group/dynamical-phase-transitions-GCAs}}.

Since the strength of the reweighting $\lambda$ which we fix during the iteration of the fixed point, acts only as the Lagrangian multiplier, it has no immediate correspondence to the value of the constraint (e.g. $\rhoinit$).
To find the concrete value, we use that at a fixed point $\chi^\to$ it holds that
\begin{align}
      \frac{\partial \Phi_{BP}(\lambda)}{\partial \lambda} = -\frac{1}{n}\left\langle \tilde{\Xi} \right\rangle_{BP} =
     - \frac{{\scriptstyle \sum_{\tx, \ty}}  \tilde{\Xi}(\tx) {\scriptstyle e^{-\lambda \tilde{\Xi(\tx)}} \chi_{\ty,\tx}^{\to} \chi_{\tx,\ty}^{\to}}}{{\scriptstyle \sum_{\tx, \ty} e^{-\lambda \tilde{\Xi}(\tx)} \chi_{\ty,\tx}^{\to} \chi_{\tx,\ty}^{\to}}}.
\end{align}
We can measure the activity $\alpha$ or the density in the attractor $\rhoattr$ by adjusting the function $\tilde{\Xi}$ correctly.
This allows us to obtain their marginals even when we did not reweight the distribution, as this corresponds to the setting where the corresponding $\lambda=0$.

Notice that the assumption of all the neighbourhoods being described by eq.~(\ref{eq:BP-equation}) is equivalent to the replica symmetric assumption which in turn on random regular graphs without another source of disorder is equivalent to the annealed calculation of the free entropy. In the present systems, the annealed calculation is non-trivial, see e.g. \cite{dandi2023maximally} and writing the BP equations~(\ref{eq:BP-equation}) is the most efficient way to obtain it we know of.

\paragraph{Application to \rebel GCAs.} 
Notice that above we wrote the equations for dynamical systems that are updated in parallel, are deterministic and run in discrete time. The update function does not distinguish between particular neighbours of a node and the connectivity graph of the neighbouring nodes is locally tree-like in the large size limit.
Finally, the size of the system's attractors has to stay constant as the system's size increases.
Since from our definitions and our empirical observations all these properties hold for the \rebel rules on random regular graphs, the (B)DCM is perfectly suitable for an analysis of the \rbl rules.

Recall that we want to answer ``What are the average properties of the typical backtracking attractor for a fixed $\rhoinit$ when $p\to\infty$?''.
One can take two approaches to this question, either by answering it starting from the initial or final configuration of the backtracking attractor.

To answer ``What are the properties later in the dynamics given that the starting point is fixed?'', we use the DCM.
This means setting $c=0$ in the backtracking attractors, we are only looking at paths.
As we increase the trajectory length $p$ we can observe how the density on the last configuration $\rho_p=\rho(\xx^p)$ evolves.

To answer ``How large is the basin of attraction of a specific type of attractor?'', we use the BDCM.
We can fix properties of the attractor, e.g. $c=2$ and $\alpha=0.5$ to identify a specific partially rattling attractor, and then increase $p$ to measure the evolution of the size of the basin of attraction in terms of its entropy density.
As one increases the length of the incoming path $p$, the analysis incorporates a growing fraction of the attractors' basin.
Comparing this entropy between different types of attractors allows us to determine which is the most numerous and typical behaviour that is observed in the large $n$ limit.

We will use these two general principles to identify analytically the dynamical phase transitions we empirically observed in Section~\ref{sec:rebel-rules-intro}.

\paragraph{Limitations and Alternative Methods.}
A significant limitation of the (B)DCM is that solving the previously mentioned fixed point equations numerically requires a computational budget which grows exponentially in $d(p+c)$ when considering a $d$-regular graph.
While the dependence on $d$ can be alleviated via dynamical programming~\cite{torrisiUncoveringNonequilibriumStationary2022}, it is prohibitive to analyse very long paths $p$ or large cycles $c$.
This means that applying the method directly is only possible for small dynamic motifs which yield interesting results only for rapidly relaxing properties at the start or end of the dynamics.
However, this is exactly what we observe for the \rebel rules and which makes the analysis with the (B)DCM feasible.

It is worth noting that by making additional assumptions, such as the one-time approximation, longer dynamics become amenable to the method.
However, this is at the cost of further uncontrolled approximations
\cite{aurellDynamicMeanfieldCavity2012,delferraroDynamicMessagepassingApproach2015,barthelMatrixProductAlgorithm2018}.
Alternative methods of analysis from statistical physics give results for simpler dynamics; examples include but are not limited to the random functions in RBNs~\cite{random_networks_of_automata, phase_transitions_in_random_networks, length_of_state_cycles_of_random_boolean_networks} or unidirectional dynamics with absorbing states~\cite{altarelliOptimizingSpreadDynamics2013,lokhovDynamicMessagepassingEquations2015}.
Another helpful feature is the relaxation of the topology, for example oriented graphs~\cite{neriCavityApproachParallel2009}, graphs with asymmetrically weighted edges~\cite{mimuraParallelDynamicsDisordered2009} for straightforward use with the DCM or independently re-sampled neighbourhoods at every iteration~\cite{vieiraThresholdVoterModel2018,phase_transition_in_the_majority_model} which are amenable to mean field methods.
However, to the best of our knowledge the (B)DCM as we use it comes closest to the very difficult case of understanding cellular automata with its rigid and deterministic architecture.

\section{Dynamical Phase Transitions for Conforming Non-Conformist GCAs}\label{sec:phase transitions}

In the following, we detail how we apply the DCM and BDCM to the examples we investigated empirically in Section~\ref{sec:rebel-rules-intro},  Fig.~\ref{fig:overview-empirics}.
Recall that for all GCAs seen previously, when $\rhoinit$ is close enough to either $0$ or $1$, the dynamics rapidly falls into one of the homogeneous attractors, while the region in between exhibits more complex dynamics. This region differs for every rule type.
The goal is to analytically identify these phase transitions between the regions precisely.
Some of these results have previously been used to demonstrate the BDCM in \cite{behrens2023backtracking}.

\paragraph{Anti-conformist GCA: 001011.}

\begin{figure*}[ht]
   \centering
   \includegraphics[width=0.31\linewidth]{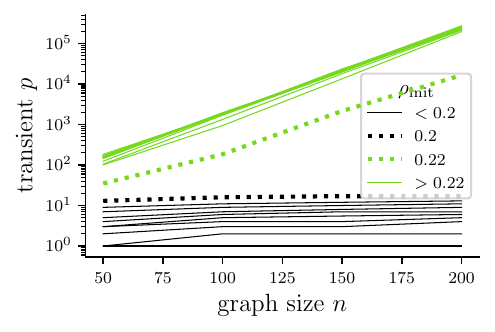}
   \includegraphics[width=0.34\textwidth]{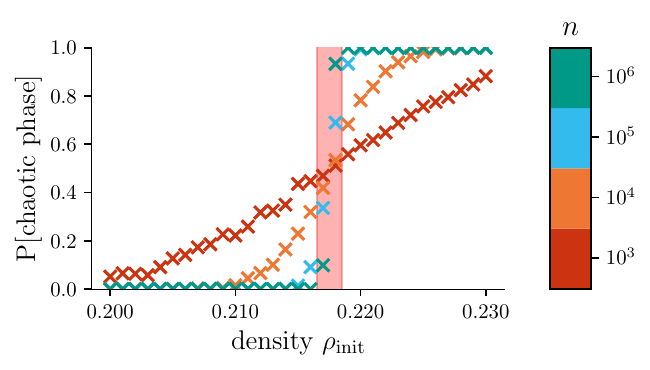}
   \includegraphics[width=0.31\linewidth]{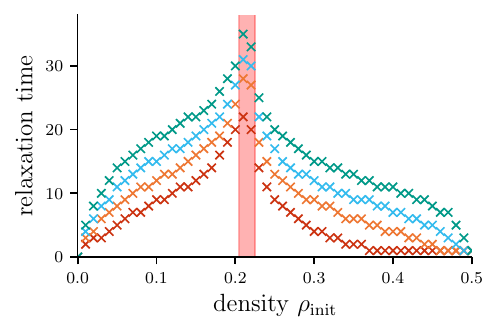}
  \caption{\textbf{Transient growth, chaotic phase classification and relaxation time for the anti-conformist GCA $001011$.} \textit{(Left)} For $\rhoinit < 0.5$ we display the transient growth for graphs of size $n\in\{50,100,150,200\}$, generated as in Fig.~\ref{fig:overview-empirics}. The resolution of $\rhoinit$ is limited by $n=50$, a stepsize of $0.02$.
  A transition between an exponential (straight line in the log-linear plot) and a much slower transient growth between $\rhoinit=0.2$ and $0.22$ is clearly visible.
  \textit{(Middle)} Empirical phase transition for the onset of a chaotic phase, which in this case is defined as the attractor taking more (chaotic) or less (homogeneous stable) than $\log(n)*100$ time steps to reach an attractor. The resolution of $\rhoinit$ is $0.001$ and narrows the interval of the dynamical phase transition down to $[0.2165,0.2185]$, the interval for $n=10^6$ between which no samples out of 1024 exhibit a behaviour that is not consistent with their phase.
  \textit{(Right)} 
  The relaxation time describes the number of time steps required until either the chaotic regime or an attractor is reached.
  We empirically conclude the system is in the chaotic regime if the densities of 100 consecutive configurations remain in the interval $(0.5-\frac{3}{\sqrt{n}}, 0.5+\frac{3}{\sqrt{n}})$.
  For all values of $n$ the maximal length of the two largest $\rhoinit$ we observe are $\rhoinit=0.21$ and $0.22$.}
   \label{fig:001011-Relaxation}
\end{figure*}

\begin{figure*}
   \centering
   \begin{minipage}{0.31\textwidth}
       \centering
   \includegraphics[width=\textwidth]{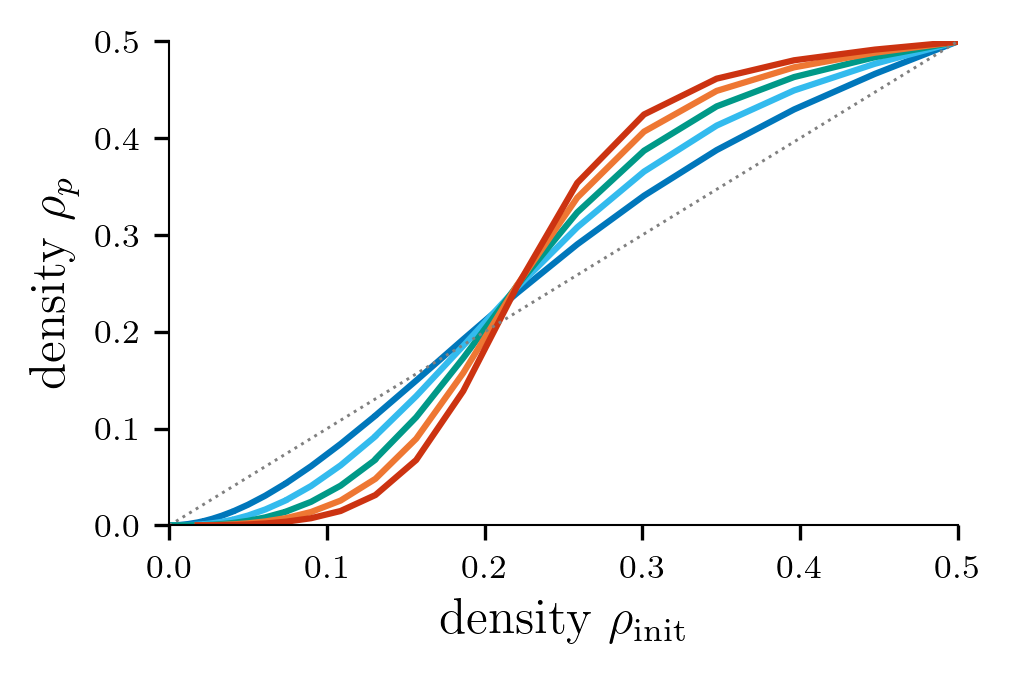}
   \end{minipage}
   \begin{minipage}{0.31\textwidth}
       \centering
   \includegraphics[width=\textwidth]{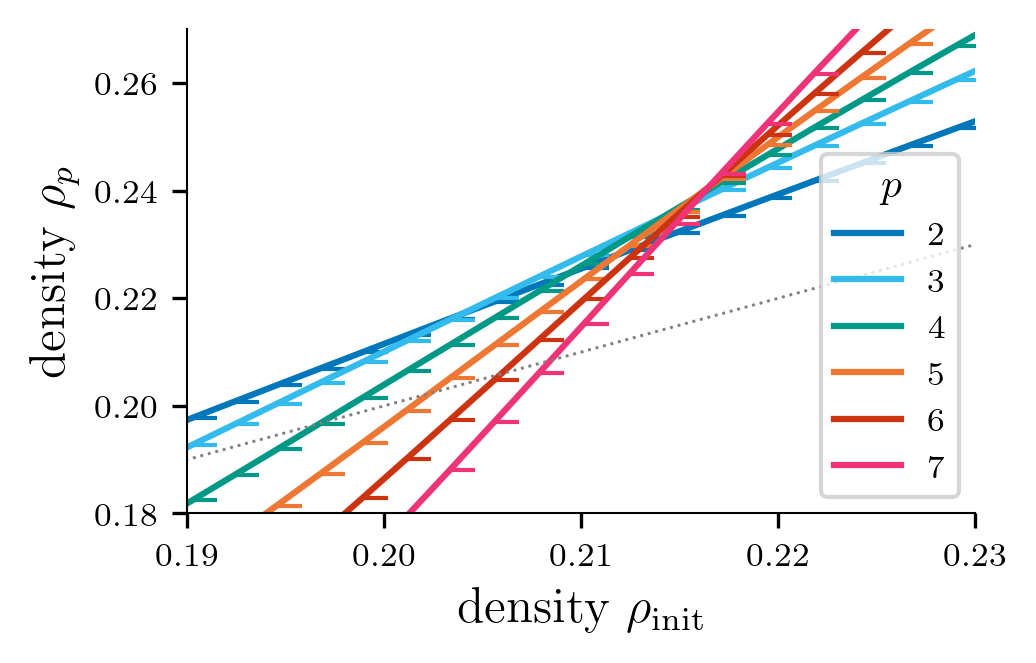}
   \end{minipage}
   \begin{minipage}{0.31\textwidth}
       \centering
   \begin{tabularx}{\textwidth}{ll}
       \toprule
        & DCM prediction \\
       $p$ & intersection of $\rho_p$ and $\rho_{p+1}$ \\
       \midrule
       2 & 0.2039 \\
       3 & 0.2142 \\
       4 & 0.2158 \\
       5 & 0.2165 \\
       6 & 0.2167 \\
       \midrule
       $\to \infty$ & $0.2168\pm0.0001$ \\
       empiric & $0.2175\pm0.001$ \\
       \bottomrule
   \end{tabularx}
   \end{minipage}
   \caption{\textbf{DCM prediction of dynamical phase transition for the anti-conformist GCA $001011$.}
   \textit{(Left)} The prediction of the DCM for the density $\rho_p$ after $p$ steps, for different initial configurations $\rhoinit$.
   \textit{(Middle)} Zoom into the region of the phase transition, with data for $p=7$ added.
   \textit{(Right)} Table of the crossover points between the different lines. The curves in the middle zoom were fitted with a linear regression and then the intersection was computed.
   Extrapolating $p\to\infty$ gives a transition at $\rhoinit=0.2168\pm0.0001$ (see Appendix Fig.~\ref{app:fig:001011-dcm-extrapolation}).
   }
   \label{fig:001011-DCM}
\end{figure*}

Recall that the anti-conformist GCA $001011$ exhibits both chaotic and ordered behaviour for different values of $\rhoinit$, but always converges to the all-1 or all-0 attractor eventually.
The dynamics of this GCA is fully deterministic, yet the configurations of trajectories in the chaotic phase look random with respect to the density $\rho$ (see e.g. Fig.~\ref{fig:example-rules}), hence the name.
The difference in behaviour between the chaotic and ordered phase clearly shows in Fig.~\ref{fig:001011-Relaxation} (left), where the transient length grows exponentially in the graph size $n$ for $\rhoinit\geq0.22$.
However, running larger system sizes than $n=200$ until convergence is prohibitively expensive, so the resolution of the transition we can obtain from this method is limited.\\
Therefore, as an additional criterion for identifying the chaotic phase for anti-conformist GCA, we check when the convergence time exceeds a threshold of $100 * \log_2(n)$.
At this point, the simulation is stopped and trajectories that have not yet converged are classified as chaotic.
Even though this heuristic is robust to changes of the factor $100$ to $50$ or $1000$, we confirm the results with another method.

Inspecting the trajectories of the density $\rho$ in Fig.~\ref{fig:example-rules}, we observe that the density of configurations in a chaotic phase is oscillating around $\rho=0.5$; more precisely it seems to remain in the interval of densities $(0.5-\frac{c}{\sqrt{n}}, 0.5+\frac{c}{\sqrt{n}})$ where $n$ is the system size and $c$ is a constant (see Appendix~\ref{app:sec:supp-material-phase transitions} for details). We use this observation as a heuristic criterion for assessing whether a trajectory has entered the chaotic phase: once a trajectory's densities stay in $(0.5-\frac{3}{\sqrt{n}}, 0.5+\frac{3}{\sqrt{n}})$ for a sufficient amount of time (100 time-steps), we conclude the trajectory is in the chaotic phase.
The time it takes to either reach this chaotic phase or an attractor is shown in Fig.~\ref{fig:001011-Relaxation} (right), it peaks around the approximate location of the dynamical phase transition.

With these three numerical experiments from Fig.~\ref{fig:001011-Relaxation}, we have a good agreement to identify a phase transition to be between $\rhoinit=0.217$ and $0.218$.
We proceed by obtaining it analytically using the DCM.

Recall that the DCM is limited to small lengths $p$ of the trajectory for which we can solve the fixed point iterations efficiently.
The question is then, how can we distinguish whether the dynamics converges fast or slow when we can look ahead only a finite number of steps $p$?

To answer this, observe that the relaxation time is extremely fast for any $\rhoinit$, even for the ones that go on to stay in the chaotic region for an exponentially long time.
Further, we observed that on average, during the chaotic phase, the density is $0.5$.
The appropriate question is then:
after $p$ steps of the DCM, what is the density of the last configuration $\rho_p$ in the large $n$ limit?
At the inflection point for growing $p$, we expect to find the dynamical phase transition.
In Fig.~\ref{fig:001011-DCM}, we show an overview and zoom-in for the relationship between $\rhoinit$ and $\rho_p$, for $p$ up to $7$.
We compare the extrapolated value for $p\to\infty$, assuming exponential convergence (see Appendix~\ref{app:sec:supp-material-phase transitions}), and our empirical extrapolation.
Indeed, the correspondence between theory and empirics is very good, with a theoretically predicted transition around $\rhoinit\sim0.2168$.

Since both the chaotic and ordered dynamics for the anti-conformist GCA $001011$ have attractors of the same type, the backtracking approach of the BDCM is not very insightful for this specific transition. However, it is useful to inspect the other CNC rules in the following.

\paragraph{Absolute Majority GCA: $0011$.}

\begin{figure*}
   \centering
   \begin{minipage}{0.31\textwidth}
   \includegraphics[width=\linewidth]{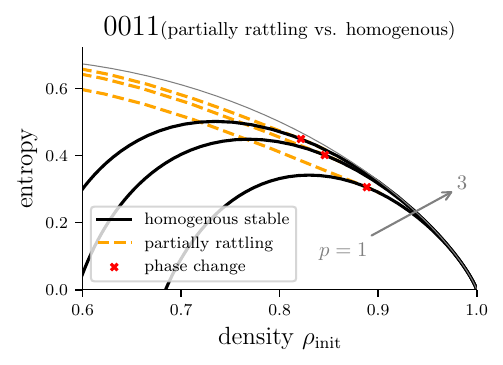}  
   \end{minipage}
   \begin{minipage}{0.31\textwidth}
   \includegraphics[width=\linewidth]{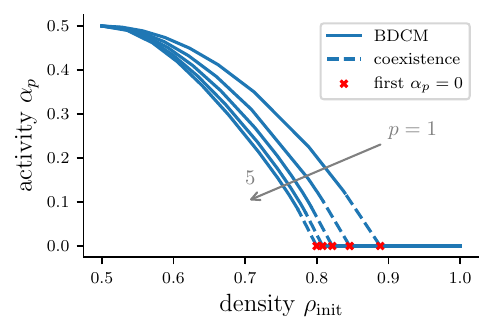}
   \end{minipage}
   \begin{minipage}{0.31\textwidth}
       \centering
   \begin{tabularx}{\textwidth}{lll}
       \toprule
        \multicolumn{3}{l}{BDCM prediction $0011$} \\
        \multicolumn{3}{l}{for first $\alpha(\rhoinit^{*,p})=0$} \\\midrule
       $p$ & $\rhoinit^{*,p}$ & $s_p^* / H(\rhoinit^{*,p})$ \\
       \midrule
       1 & 0.8885 & 0.873\\
       2 & 0.8459 &  0.933\\
       3 & 0.8216 & 0.958\\
       4 & 0.8080 & 0.971\\
       5 & 0.7996 & 0.979\\
       \midrule
       $\to \infty$ & \multicolumn{2}{l}{$0.7875\pm0.005$} \\
       empiric & \multicolumn{2}{l}{$0.785\pm0.005$} \\
       \cite{kanoriaMajorityDynamicsTrees2011} & $>0.7865$ \\
       \bottomrule
   \end{tabularx}
   \end{minipage}
   \caption{\textbf{BDCM prediction for the absolute majority GCA $0011$.} \textit{(Left)} Entropy of the basin of attraction for the homogeneous and partially rattling attractors, for increasing path lengths $p=1,2,3$. \textit{(Middle)} The activity in the limit cycle for fixed points for $(p/c=2)$ backtracking attractors. The dashed line shows the range of $\rhoinit$ for which our numerics did not find any fixed points.  \textit{(Right)} The table shows the values of the smallest $\rhoinit^{*,p}>0.5$ for which $\alpha(\rhoinit^{*,p})=0.0$ together with the normalized entropy at the corresponding given $\rhoinit$. It shows the extrapolation $p \to \infty$ and compares it with the numerical results and related work (see Appendix~\ref{app:sec:supp-material-phase transitions}).}
   
   \label{fig:0011-bdcm}
\end{figure*}

For the absolute majority GCA $0011$, the convergence is logarithmic independently of $\rhoinit$ and the system's phases differ only in the type of attractor they converge to.

In Fig.~\ref{fig:0011-bdcm} we show the entropy of backtracking attractors with a path length $p=1,2,3$ obtained via the BDCM.
Here, the entropy represents the size of its basin of attraction when stepping back $p$ steps from the attractor, for a specific $\rhoinit$.
Each differently styled line represents a single type of attractor.
Their entropy was obtained by solving the BDCM fixed point iteration under the constraint matching the respective attractor properties, i.e. $c=1,2$.
In addition, the value of $\rhoinit$ was constrained, giving the final result.
In the large $n$ limit only the types of attractors with the maximum entropy are expressed.
Therefore, the correct way to interpret the plots is to check which attractor type has the maximum entropy for every $\rhoinit$ --- this phase will be the one which is typically observed in large systems.

At $p=0$, we would only count the attractors, without their basin, essentially using the method from \cite{hwangNumberLimitCycles2020}.
However, only as we increase $p$ and incorporate the basin of attraction, we observe that the overall picture from the empirics Fig.~\ref{fig:overview-empirics} is reproduced qualitatively by the BDCM: For large $\rhoinit$, it shows the all-one attractor.
Decreasing $\rhoinit$ around 0.5, one finds the partially rattling attractor.

Since a $(p/c=1)$ backtracking attractor is also a $(p/c=2)$ backtracking attractor, the two entropy curves naturally merge when the $(p/c=2)$ backtracking attractors reduce to attractors that are of length $c=1$. This merge between the two curves is the dynamical phase transition at a given fixed $p$, see Fig.~\ref{fig:0011-bdcm} (left).
Inspecting the fixed point for $c=2$, we indeed find for large enough $\rhoinit$ that the activity $\alpha$, the fraction of rattling nodes in the limit cycle, becomes essentially zero (Fig.~\ref{fig:0011-bdcm} (middle)). This indicates that the number of such rattling nodes no longer scales in $O(n)$ and that the fixed point only considers limit cycles of length $c=1$.
Recording the switch from $\alpha=0$ to $\alpha>0$ gives the dynamical phase transition, as shown in the Table on the right in Fig.\ref{fig:0011-bdcm}.
Even though we did not compute values larger than $p=5$, we extrapolate the BDCM result to $p\to\infty$ to make our theoretical prediction.
This agrees well with the empirical prediction (Appendix~\ref{app:sec:supp-material-phase transitions}).

\paragraph{Stubborn Independent GCA: $00$$+$$11$.}
\begin{figure*}
\centering
   \begin{minipage}{0.31\textwidth}
   \includegraphics[width=\linewidth]{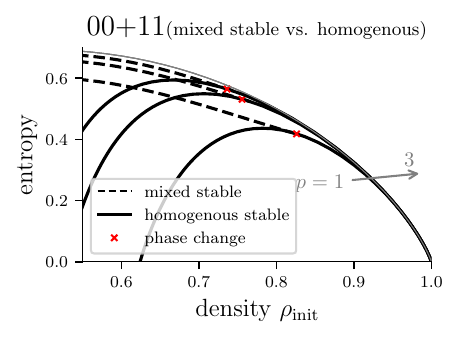}  
   \end{minipage}
   \begin{minipage}{0.31\textwidth}
   \includegraphics[width=\linewidth]{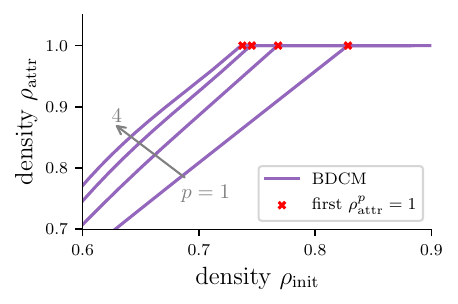}
   \end{minipage}
   \begin{minipage}{0.31\textwidth}
       \centering
   \begin{tabularx}{\textwidth}{lll}
       \toprule
        \multicolumn{3}{l}{
         BDCM prediction $00$$+$$11$} \\
         \multicolumn{3}{l}{
         for first $\rhoattr(\rhoinit^{*,p})=1$} \\\midrule
       $p$ & $\rhoinit^{*,p}$ & $s^*_p / H(\rhoinit^{*,p})$ \\
       \midrule
       1 & 0.828 & 0.909 \\
       2 & 0.768 & 0.965\\
       3 & 0.745& 0.983 \\
       4 & 0.737& 0.990 \\
       5 & 0.733&0.994 \\
       \midrule
       $\to \infty$ & $0.73\pm0.005$ \\
       empiric \cite{behrens2023backtracking} & $0.731\pm0.002$ \\
       \bottomrule
   \end{tabularx}
   \end{minipage}
   \centering
   \caption{\textbf{BDCM prediction for the stubborn independent rule 00+11.}
   \textit{(Left)} Entropy of the basin of attraction for the all-1 and mixed stable attractors respectively, for increasing steps into the basin of attraction $p=1,2,3$. The dynamical phase transition is marked in red. \textit{(Middle)} The density of 1's in the attractor, $\rhoattr$, as a function of $\rhoinit$ for attractors with $c=1$.  
   \textit{(Right)} The transitions is the first $\rhoinit$ for which the $\rhoattr=1$. These values are recorded in the table together with the entropy of the basin of attraction at that point.
   The extrapolation agrees well with the empirical estimate of the transition from the maximal slowing down (see Appendix~\ref{app:sec:supp-material-phase transitions}).}
   \label{fig:00+11-bdcm}
\end{figure*}
We can do a similar type of analysis for the stubborn independent GCA $00$$+$$11$. This is the GCA where the node in the weak agreement region is stubborn, i.e. it sticks with its own opinion. In this analysis, we distinguish between two types of attractors that go either to the homogeneous all-1 or mixed stable state, which both have a limit cycle length of $c=1$.
Recall that the mixed stable state is defined to be an attractor where the density in the attractor $\rhoattr$ is not 0 or 1 (see Table~\ref{tab:attractor-types}).
In Fig.~\ref{fig:00+11-bdcm} (left) we show the entropy for small $p$ in terms of $\rhoinit$.
To identify the dynamical phase transition we again track the spot where the attractor type with the maximum entropy switches over.
Here, this is a merge of the two curves again.
While we can restrict the fixed point iteration to variables that always end up in the all-1 attractor, for mixed stable attractors there is so far no technical means of constraining it to a non-zero $\rhoattr$.
This is why we track the $\rhoattr$ as a function of $\rhoinit$ in Fig.~\ref{fig:00+11-bdcm} (middle) and record when this property becomes close enough to 1.0, giving us the value of change.
The table on the right of Fig.~\ref{fig:00+11-bdcm} records these values.
The extrapolation to $p\to\infty$ matches well with the empirical result.

\paragraph{Volatile Independent GCA: $00$$-$$11$.}

\begin{figure*}
   \centering
   \includegraphics[width=\linewidth]{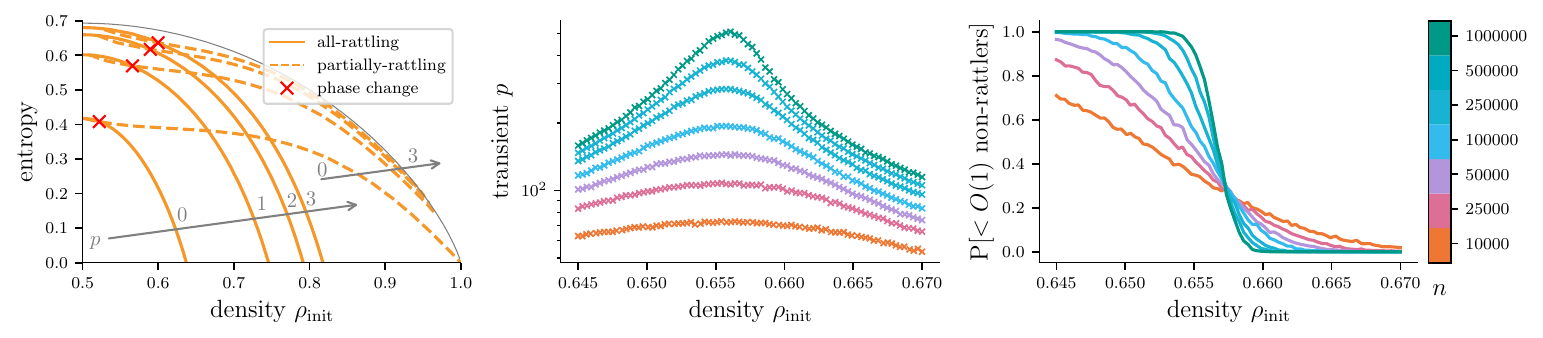}
   \caption{\textbf{Dynamical phase transition for the volatile independent GCA $00$$-$$11$.} Comparison of the analytical and empirical prediction of a dynamical phase transition for the volatile independent GCA $00$$-$$11$. We examine the transition between the all and partially rattling 2-cycles.  \textit{(Left)} Analytical prediction of the entropy for each $\rhoinit$ for the two different types of attractors for increasing transient lengths $p$.
   The intersection of the two entropy marks the phase transition for a given $p$ and is marked in red.
   Because the computed entropy is not close enough to the maximal entropy, as shown by the grey line, the approximation of the transition is not very conclusive and extrapolating the four data points would result in very high uncertainty.  \textit{(Middle)} Zoom in on the average transient length around the phase transition from Fig.~\ref{fig:overview-empirics}. \textit{(Right)} Probability of obtaining a smaller than $o(n)$ fraction of rattlers, i.e. the fraction of nodes in the attractor. To determine a reasonable threshold for having a constant $o(n)$ fraction of rattlers, when $n$ is finite, we analysed the scaling of the rattler fraction as a function of $n$, which resulted in an attractor having no more than 0.07\% of non-rattlers to be classified as a partially-rattling attractor (Appendix~\ref{app:sec:supp-material-phase transitions}). While the thresholds agree roughly, the accuracy is worse than for the GCAs discussed previously.}
\label{fig:00-11_bdcm}
\end{figure*}

The volatile independent GCA $00$$-$$11$ is slightly more complex, as it has more phases than the GCAs discussed previously, and four dynamical phase transitions (Fig.~\ref{fig:overview-empirics}).
Since the transition between partially rattling and homogeneous phase is similar to the GCA $0011$, we discuss only the transition between all-rattling and partially-rattling, i.e. the change between attractors of length $c=2$ where either all nodes change (activity $\alpha=1.0$) or some of them are fixed ($\alpha < 1$).

In Fig.~\ref{fig:00-11_bdcm} the entropy of the two phases is shown on the left. This time, the two fixed points intersect, and do not merge.
However, the fraction of the basin of attraction covered by the $p=4$ steps that are taken back, is smaller than in the other examples.
This can be viewed as a reason for which this dynamical transition is correct qualitatively, but the approximation is not precise.
The fact that for this GCA, more steps back are necessary, reflects the common observation that close to phase transitions the convergence time increases, which makes the use of the BDCM more challenging computationally in its vicinity by default.

\section{Conclusion and Open Questions}

In this work, we use tools from statistical physics -- the dynamical cavity method and its backtracking version -- to demonstrate that they are powerful for deriving analytical results on the global dynamics of discrete dynamical systems in the large system size limit.

Concretely, we study a class of graph cellular automata called the \rebel GCAs that can be interpreted as various models of opinion formation dynamics. We argue that such systems exhibit a rich set of dynamical phases defined by their different transient and attractor properties, and we show the existence of sharp transitions between such phases in terms of the initial configuration density.

For two specific examples with small degrees, we showed how the (B)DCM methods are applied and predict the phase transitions. We show that our analytic predictions agree well with numerical estimates for reasonably large systems.

Such results enforce the narrative that for discrete dynamical systems, different choices of initial configurations can lead to qualitatively different regimes of the system's behaviour.

\paragraph{Relationship between CAs and GCAs.}
In its formulation, the graph cellular automata are extremely close to classical cellular automata -- they only differ in how their nodes are connected.
Whereas for CAs, the connectivity network is given by a regular grid, the GCAs' connections are defined by a random regular graph. As such, deriving analytical results about their global dynamics is challenging and our work shows a variety of new results about such systems.

It is not yet clear in how far our results for the random regular graphs (GCAs) transfer to the regular lattice (CAs). Even though classical CAs are not amenable to the analysis via (B)DCM, a numerical investigation is still possible. Previous work has shown that similar types of attractors and phases do occur on the lattice \cite{bagnoliPhaseTransitionsCellular2014}, but our own preliminary empirical investigations did not show an immediate and unambiguous connection. We leave a thorough investigation of these empirics for future work.

Clearly, an analytic method capable of directly handling deterministic CAs directly rather than extrapolating behaviour from the regular GCAs or probabilistic cellular automata \cite{petersenPhaseTransitionElementary1997,bagnoliPhaseTransitionsCellular2014} is a challenging goal.

\paragraph{Limitations of the (B)DCM.} A major drawback of the DCM and BDCM is the exponential computational barrier which depends on the length of the analysed motif $p+c$. Even when the system typically relaxes fast, as previously noted, this limitation may lead to less accurate estimates of the transition \cite{behrens2023backtracking} as around phase transitions the transient length may increase due to critical slowing down.
Therefore, it seems worth investigating if and how approximations to the DCM \cite{aurellDynamicMeanfieldCavity2012,zhangInferenceKineticIsing2012,delferraroDynamicMessagepassingApproach2015,barthelMatrixProductAlgorithm2018,torrisiUncoveringNonequilibriumStationary2022,hurryDynamicsSparseBoolean2022a} would give new insights into longer time scales, and if they remain accurate around phase transitions or suffer from similar limitations.
Moreover, it is an open task to adapt such approximations to the backtracking version of the DCM.

\paragraph{Large degrees.}
For very large degrees $d$, which scale in the size of the graph $n$, we empirically extrapolate our results. We deduce from the numerics that a scaling of the weak agreement threshold $\theta$ approximately as $\sqrt{d}$ maintains the dynamical phase transitions we showed for the small degrees.
For the absolute majority and stubborn/volatile independent GCAs we conjecture based on our numerical experiments that only the behaviour that we showed previously will occur.
However, preliminary results for the anti-conformist GCAs showed that new types of behaviour emerge when we increase $d$, hinting at further dynamical phase transitions that require a higher resolution in $d$ to manifest. We leave a thorough investigation of this rule space and its peculiarities for future work.

\paragraph{Short attractors.}
While for the absolute majority and volatile/stubborn independent GCAs only short attractors of length 1 and 2 can occur \cite{golesPeriodicBehaviourGeneralized1980}, we showed that empirically the same holds true for the anti-conformist GCAs on sparse random regular graphs of large size. Based on this evidence, we conjecture that in the large $n$ limit such GCAs typically only has short attractors for finite $d$. This statement remains to be proven.

\paragraph{Phase transitions and complexity.}
There has been a plethora of works on dynamics of discrete systems that focus on their complex behaviour -- this is typically associated with intriguing visualizations of the systems' space-time diagrams or with the capacity to compute challenging tasks \cite{universality_and_complexity_in_ca, winning_ways_for_mathematical_plays, kaneko}. Many attempts at formalizing the notion of complexity have been given with the general belief that the region of complex behaviour is located at a phase transition between ``ordered'' and ``chaotic'' systems \cite{random_networks_of_automata, edge_of_chaos}.

In our work, we do not explore the phase transition in the space of systems. Rather, for a fixed GCA, we describe the phase transition in the space of its initial configurations. This transition becomes particularly interesting for the anti-conformist GCAs that, near the transition, abruptly switch from logarithmic convergence to attractors (associated with simple behaviour) to an exponential one (interpreted as chaotic behaviour) \cite{transient_classification}. As such, it becomes very interesting to ask: Is the behaviour of the system near the phase transition qualitatively different? Does it show some signs of ``complexity''? From Fig. \ref{fig:001011-Relaxation}, middle, it is apparent that as we increase the system's size, near the phase transition the system converges to its typical behaviour much more slowly than away from the transition. Thus, in our case, the complexity arises from deciding what type of behaviour the system will settle to near the transition. However, assessing the system's complexity near the transition would require carefully choosing a formal metric of complexity. Therefore, we leave such investigations for possible future work.

\paragraph{Opinion Dynamics.} As a side product, we investigated our version of a popular framework from opinion dynamics \cite{survey_on_nonstrategic_models_of_opinion_dynamics} on a sparse graph. It encompasses a local update rule that seems anecdotally ubiquitous in popular culture: The conforming anti-conformist. This is an agent who only acts in favour of the minority when this minority is not too small, i.e. when the race between the majority and minority is tight.
Our analysis showed that such behaviour allows for two opinions to co-exist for a prolonged period of time in the system and thereby maintains a diversity of opinions.

\begin{acknowledgments}
Our work was supported SVV-2020-260589, by the Czech project AI$\&$Reasoning CZ.02.1.01/0.0/0.0/15\texttt{\char`_}003/0000466 and the European Regional Development Fund, and by Grant Schemes at CU, reg. no. CZ.02.2.69/0.0/0.0/19\texttt{\char`_}073/0016935.
\end{acknowledgments}
\newpage

\newpage
\FloatBarrier
\onecolumngrid
\appendix

\section{Larger degree behaviour}\label{app:sec:large-d}
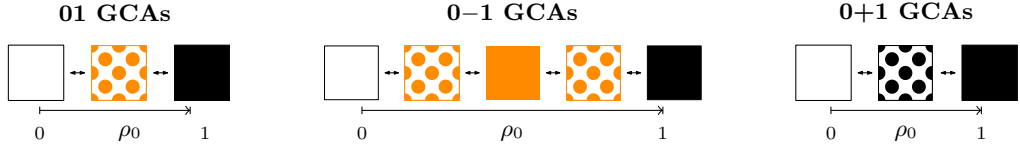
\begin{figure*}[ht]
\centering
   \begin{tikzpicture}
       \node at (-.2,1) {\textbf{01 GCAs}};
       \node at (-0.15,0.2) {\includegraphics[width=.165\linewidth]{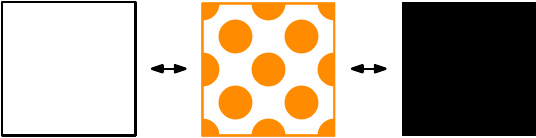}};
       \draw[->] (-1.2,-.3) -- (.8,-.3);
       \draw[-] (-1.2,-.25) -- (-1.2,-.35);
       \draw[-] (.8,-.25) -- (.8,-.35);
       \node at (-1.2, -.6) {\scriptsize $0$};
       \node at (1, -.6) {\scriptsize $1$};
       \node at (0, -.6) {$\rho_0$};
   \end{tikzpicture}
   \hspace{2.5em}
   \begin{tikzpicture}
       \node at (0,1) {\textbf{0$-$1 GCAs}};
       \node at (0,0.2) {\includegraphics[width=.28\linewidth]{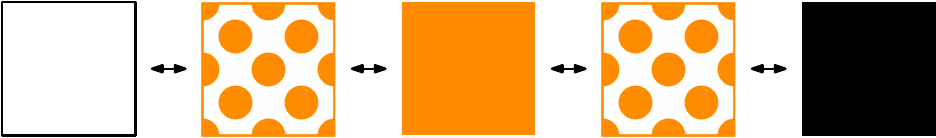}};
       \draw[->] (-2,-.3) -- (2,-.3);
       \draw[-] (-2,-.25) -- (-2,-.35);
       \draw[-] (2,-.25) -- (2,-.35);
       \node at (-2, -.6) {\scriptsize $0$};
       \node at (2, -.6) {\scriptsize $1$};
       \node at (0, -.6) {$\rho_0$};
   \end{tikzpicture}
   \hspace{2.5em}
   \begin{tikzpicture}
       \node at (0,1) {\textbf{0+1 GCAs}};
       \node at (0,0.2) {\includegraphics[width=.165\linewidth]{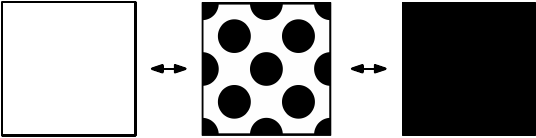}};
       \draw[->] (-1,-.3) -- (1,-.3);
       \draw[-] (-1,-.25) -- (-1,-.35);
       \draw[-] (1,-.25) -- (1,-.35);
       \node at (-1, -.6) {\scriptsize $0$};
       \node at (1, -.6) {\scriptsize $1$};
       \node at (0, -.6) {$\rho_0$};
   \end{tikzpicture}
       \caption{Phase diagram scheme for GCAs with rules of type \textbf{01}, \textbf{0$-$1} and \textbf{0+1}. While for all GCAs the homogeneous stable all-white and all-black phases are at the end of the $\rho_{\mathrm{init}}$ spectrum, the intermediate behaviour is qualitatively different. We argue that for each rule type, when increasing $\rho_{\mathrm{init}}$ from 0 to 1, the phases occurring always obey the order illustrated in the diagram, though, for degenerate cases, some of the phases might be missing (e.g., the constant 0 GCA only has the homogeneous all-0 phase).} 
   \label{fig:the-transition-overview}
\end{figure*}

We studied the dynamics of all \rebel rules for connectivity $d=3$ and $d=4$ and observed the following general trend shown in Fig.~\ref{fig:the-transition-overview}.

\begin{figure*}
   \centering
   \includegraphics[width=\textwidth]{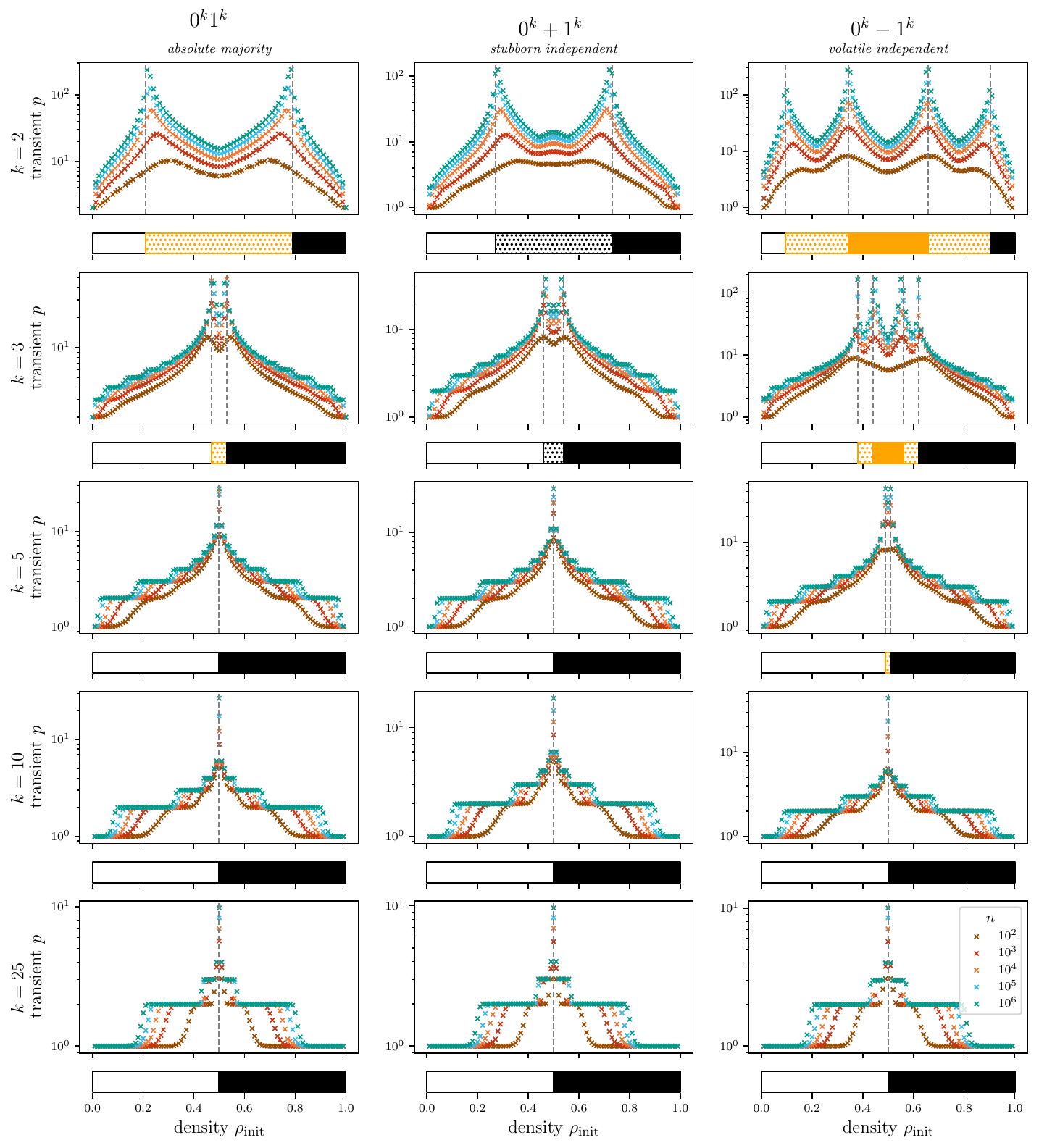}
   \caption{\textbf{Weak agreement region with $\theta \in \{0,1 \}$.} Increasing the degree $d$ for the absolute majority and stubborn/volatile independent GCAs, while keeping the weak agreement region constant. Eventually, all GCAs behave like the absolute majority GCA. Samples were obtained as described in Fig.~\ref{fig:overview-empirics}.}
   \label{app:fig:transients}
\end{figure*}

Only the volatile independent rule types \textbf{0--1} exhibit both the all-rattling and partially-rattling phases; whereas the stubborn independent rule types only exhibit stable phases.
One interesting question is: ``How does the phase transition behaviour scale for larger values of $d$?''
Fig.~\ref{app:fig:transients} illustrates that if the threshold $\theta$ remains constant as we increase the connectivity $d$, the interesting region of $\rhoinit$ shrinks and almost all initial densities exhibit fast convergence to either the all-0 or all-1 attractor. Eventually, as the degree grows all these rules behave as the majority rule.

\begin{figure*}
   \centering
   \includegraphics[width=0.62\textwidth]{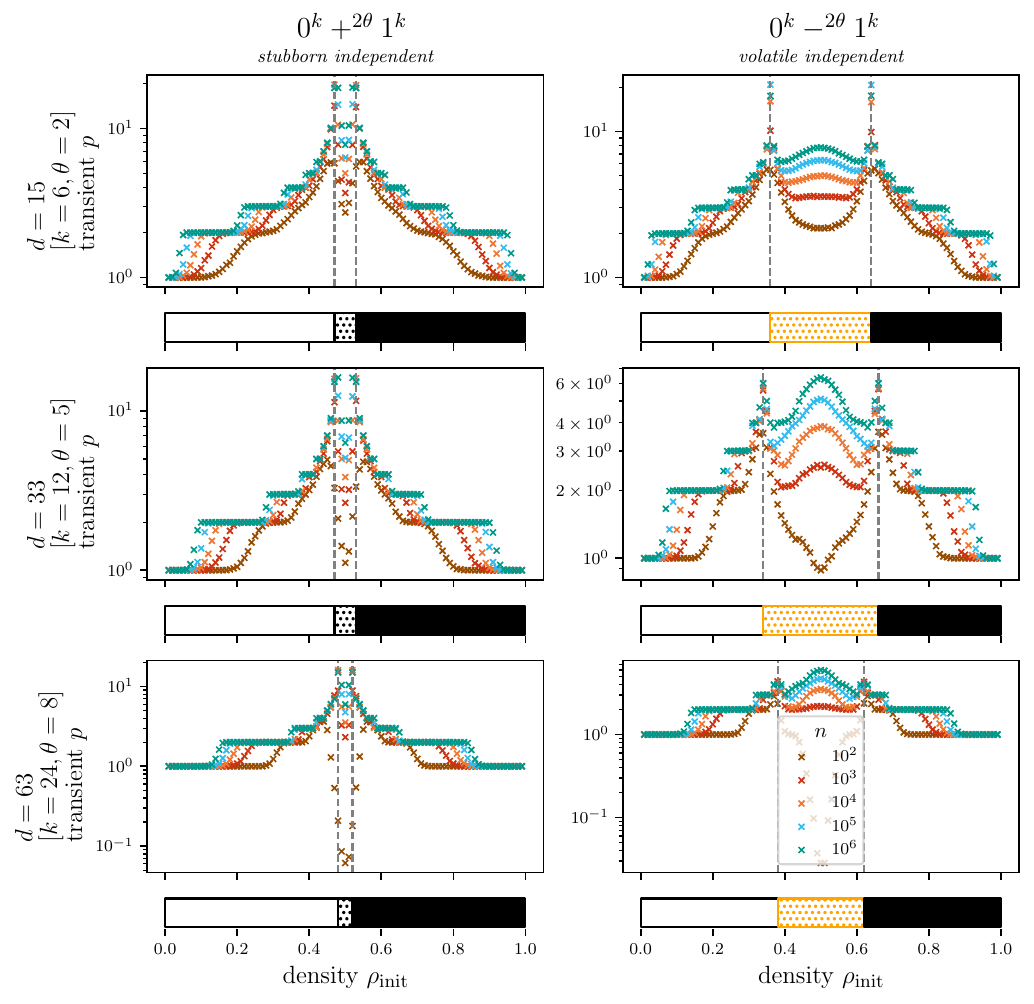}
   \includegraphics[width=0.35\textwidth]{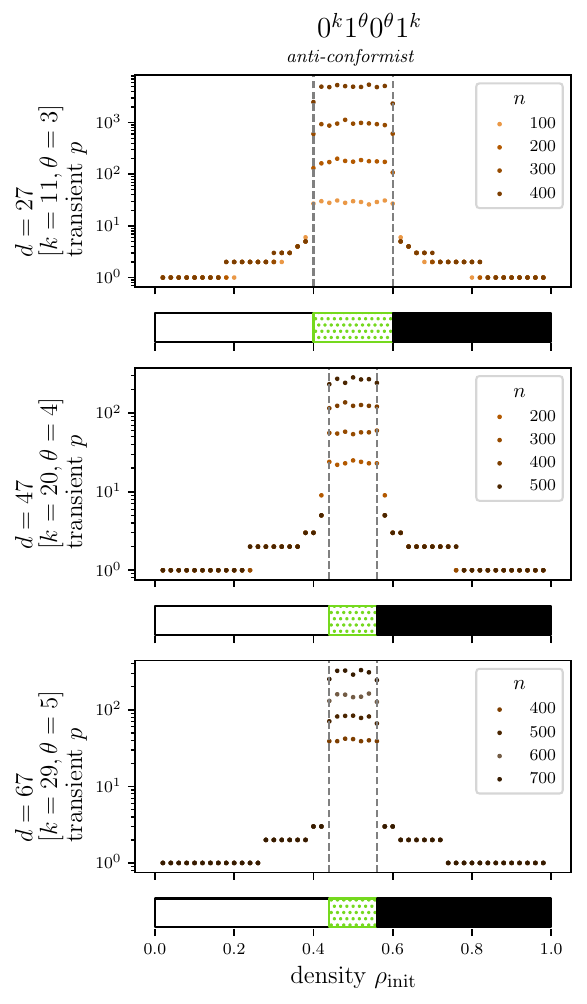}
   \caption{\textbf{Weak agreement region with scaling (very roughly) in $\theta \sim \sqrt{d}$.} Increasing the degree $d$ for the stubborn/volatile independent GCAs and the anti-conformist GCA, while scaling the weak agreement region (very roughly) as $\sqrt{d}$. Samples were obtained as described in Fig.~\ref{fig:overview-empirics}.}
   \label{app:fig:dsqrtscaling}
\end{figure*}

Let $k \in \N$ and $\theta \in \N$. We can parameterize the \rebel rules with connectivity $d$ odd in the following way:
\begin{alignat*}{3}
   \text{stubborn independent: }  {0^k +^{2\theta} 1^k}\\
   \text{volatile independent: }  {0^k -^{2\theta} 1^k}\\
   \text{anti-conformist: }  {0^k 1^\theta 0^\theta 1^k} 
\end{alignat*}
with $d = 2k + 2\theta -1$. We note that the parameter $\theta$ indeed corresponds to the threshold parameter from the definition of CNC rules in Section \ref{sec:rebel-rules-intro}. A few examples of the dynamical behaviour for \rebel rules with larger $d$ and $\theta$ are shown in Fig.~\ref{app:fig:dsqrtscaling}.
For the stubborn/volatile independent GCAs we observe that if $\theta$ scales approximately as $\sqrt{d}$, the phase transitions are preserved (Fig.~\ref{app:fig:dsqrtscaling}, left). Once $\theta$ deviates from $\sqrt{d}$ the transitions may collapse and certain phases are no longer present (Fig.~\ref{app:fig:transients}). The situation, however, looks more complicated for the anti-conformist GCAs. In Fig.~\ref{app:fig:dsqrtscaling}, right, we picked very specific values of $k$ and $\theta$, for which the general phase transition trend with the interesting region of alternate behaviour around $\rhoinit=0.5$ is present. We highlight that different values of $k$ and $\theta$ yielded new types of behaviour for the anti-conformist GCAs that need further investigation and are left for future work.

\FloatBarrier

\section{Supporting Empirics for Phase Characterization}\label{app:sec:supporting_numerics_phasecharacterization}

\begin{figure*}[ht]
   \centering
   \includegraphics[width=0.4\textwidth]{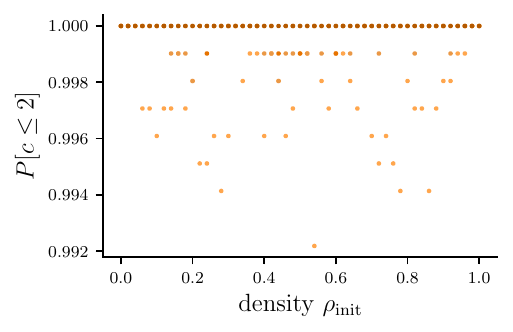}
   \includegraphics[width=0.4\textwidth]{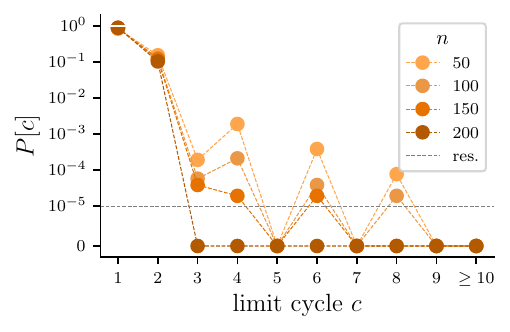}
   \caption{\textbf{Length of limit cycles for the anti-conformist GCA (code 001011).} We show the length of the limit cycles for the data collected for Fig.~\ref{fig:overview-empirics}, i.e. $4$-regular graphs and different initial densities. \textit{(Left)} Probability that a sample has a limit cycles length $c\leq2$. Since we only sample few such long limit cycles, we combine the data for the different initializations $\rhoinit$ on the \textit{(right)}. The dashed grey line shows the minimal resolution we are limited to due to our sample size, which was 1024 for each of the 100 different initial densities.}
   \label{app:fig:cycle-lengths}
\end{figure*}
In Fig.~\ref{app:fig:cycle-lengths} we investigate the lengths $c$ of the limit cycles for the anti-conformist GCA (code 001011).
There, we almost always sample short attractors.
As $n$ grows, the number of large limit cycles drops rapidly.
This leads us to the conjecture, that in the large $n$ limit, for the anti-conformist GCAs, any attractors with sizes $c>2$ will become vanishingly unlikely.

\newpage
\FloatBarrier
\section{Supporting Material for Dynamical Phase Transition Predictions using the (B)DCM and Empirical Methods}\label{app:sec:supp-material-phase transitions}

\begin{figure*}[ht]
   \centering
   \includegraphics[width=0.4\linewidth]{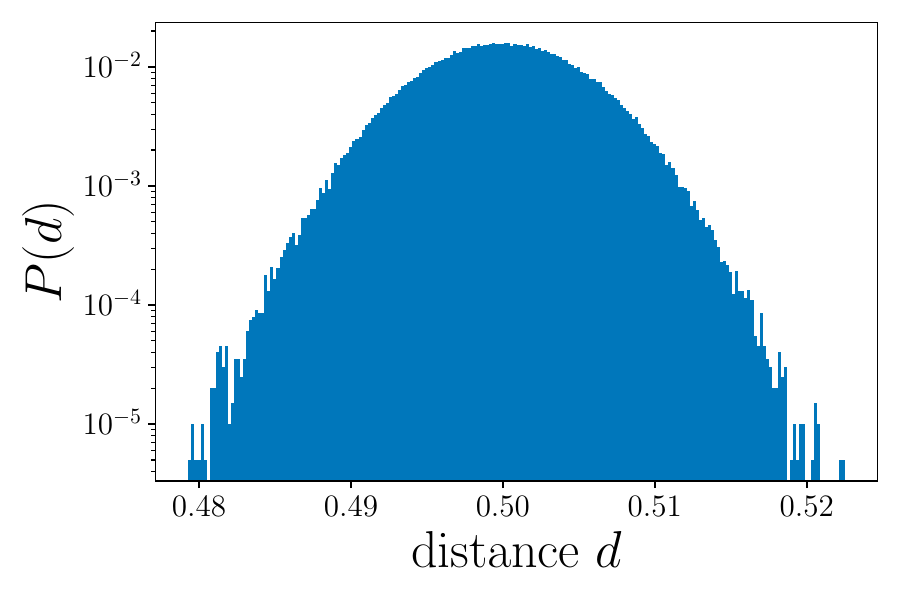}
   \caption{\textbf{Illustration of chaotic phase for anti-conformist GCA $001011$: evolution of distances of two close-by trajectories.} For GCA $001011$ we randomly generated an initial configuration with length $n=10000$, and density $\rhoinit = 0.5$ and a close-by configuration with $\epsilon \cdot n$ different bits; $\epsilon=0.01$. While both configurations' trajectories remain in the chaotic regime, we measure their average Hamming distance and plot the probabilities for 100 such experiments averaged.}
\label{app:fig:001011_distance_evolution}
\end{figure*}

In Fig.~\ref{app:fig:001011_distance_evolution} we show yet another property of the chaotic phase of anti-conformist GCAs. The graph suggests the evolution of distances of two close-by initial configurations follows a (pseudo) random walk.

\begin{figure*}[ht]
   \centering
   \includegraphics[width=0.4\linewidth]{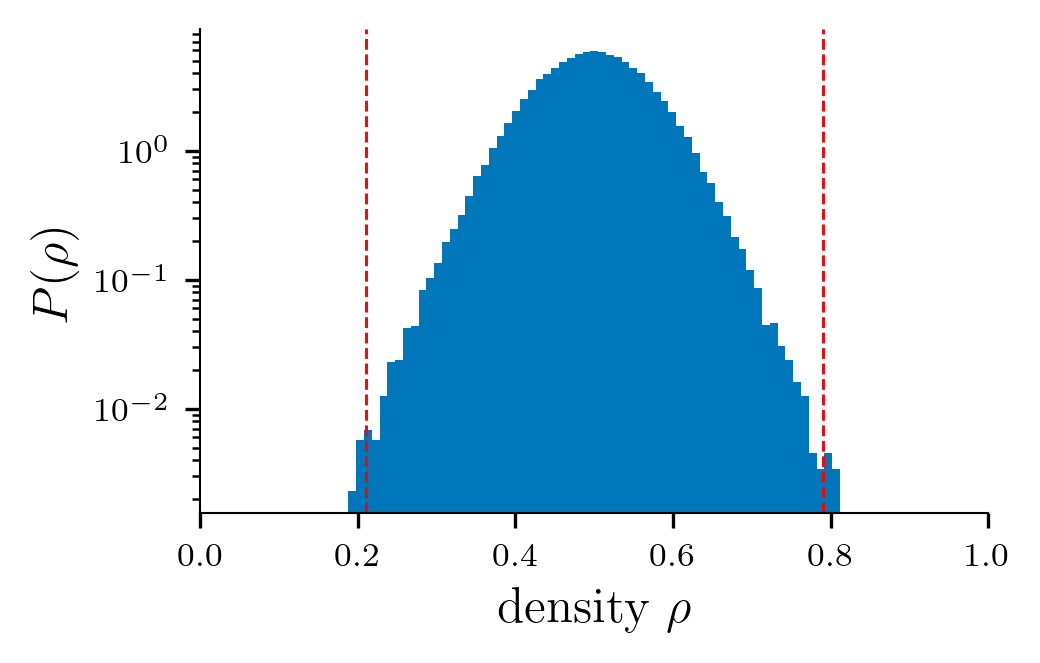}
   \caption{
    \textbf{Histogram of the density $\rho$ for transients with a chaotic behaviour (after relaxation and before convergence to an attractor) for the anti-conformist GCA $001011$.} We sample 1024 graphs of size $n=100$ with the dynamics of the rule $001011$ started at $\rhoinit=0.2$. The first 100 time steps after the start of the dynamics and last 100 time steps before reaching the attractor are removed for every sampled trajectory.
    Since the convergence is fast in the phase where an attractor is reached rapidly (see Fig.\ref{fig:overview-empirics}), such a cut-off effectively removes all transients that converge rapidly (less than 200 time steps long).
    The histogram shows the density of the remaining trajectory which exhibits a chaotic behaviour.
    The red lines mark the phase transition measured empirically for rule $001011$ between the phase of rapid convergence to the homogeneous state (left and right side) and the chaotic phase (between the red lines). Note that during this time almost all samples lie within the regime of the chaotic phase.}
\label{app:fig:001011_density_evolution}
\end{figure*}

In Fig.~\ref{app:fig:001011_density_evolution} we demonstrate that the value of the densities observed during the chaotic phase to an attractor, fall into the interval between the dynamical transition lines.

\begin{figure}
   \centering
   \includegraphics[width=0.4\textwidth]{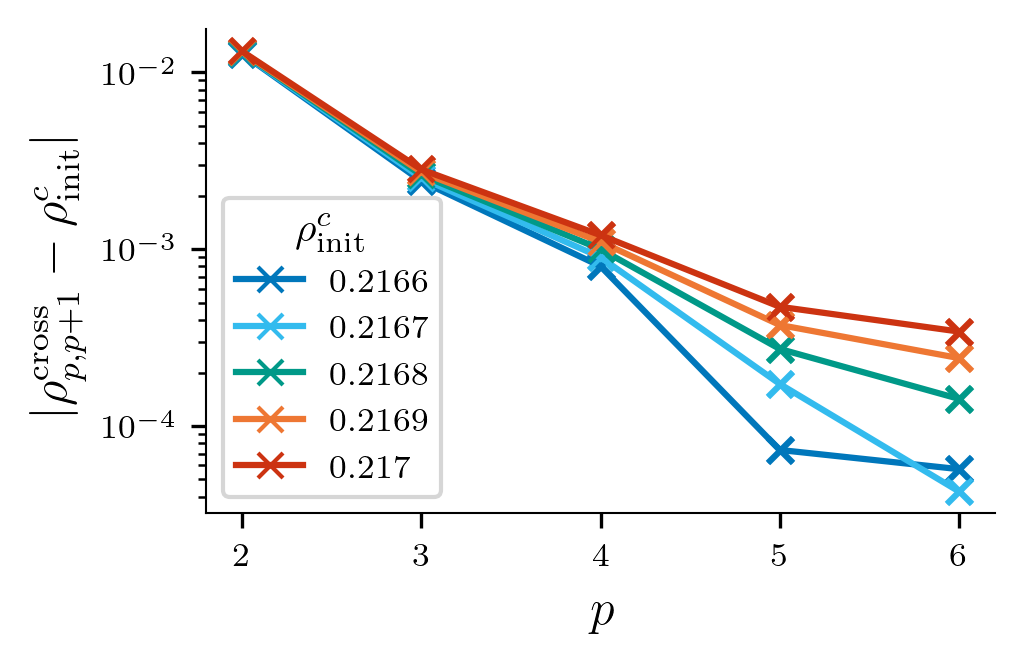}
   
   \caption{\textbf{Anti-conformist GCA $001011$ -- Extrapolation.} Extrapolation of the crossover points from Fig.~\ref{fig:001011-DCM} to $p\to\infty$. The plot shows the distance to the critical $\rhoinit^c$, the possible location of phase transition points. Under the assumption of an exponentially fast convergence in $p$, the best fit seems to lie at roughly $\rhoinit^c = 0.2168\pm0.0001$.}
   \label{app:fig:001011-dcm-extrapolation}
\end{figure}

In Fig.~\ref{app:fig:001011-dcm-extrapolation} we show how we extrapolated the $p=\infty$ behaviour from the first $7$ time steps using the DCM for the anti-conformist GCA $001011$.

We do the same for the absolute majority GCA $0011$, in addition to precise numerics in Fig.~\ref{app:fig:001011-dcm-extrapolation}.

Similarly, the BDCM results for the stubborn independent GCA $00$$+$$11$ are extrapolated in Fig.~\ref{app:fig:00+11-supplements}.

\begin{figure*}[ht]
   \centering
   \includegraphics[width=0.4\linewidth]{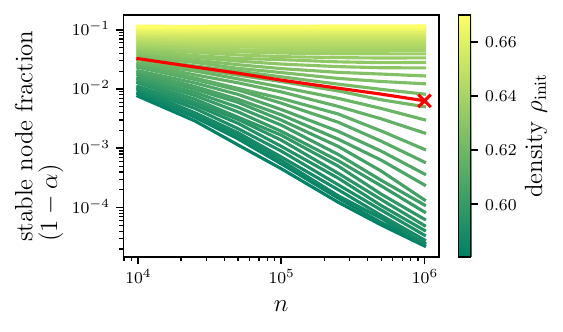}
   \caption{\textbf{Scaling of the non-rattling nodes for different $\rhoinit$.} For some $\rhoinit$ the non-rattling nodes scales logarithmic in $o(n)$, for others they are a constant fraction in $\Theta(n)$. The red value marks the intermediate threshold we selected to classify the all-rattling and partially rattling phases in Fig.~\ref{fig:00-11_bdcm} (right), which had 0.7\% of stable nodes.}
\label{app:fig:00-11_threshold_scaling}
\end{figure*}

In Fig.~\ref{app:fig:00-11_threshold_scaling} we base our selection of the threshold to distinguish the all and partially rattling phase on numerical evidence that shows how the scaling of the activity $\alpha$ behaves differently for each $\rhoinit$.

\begin{figure*}
   \centering
   \includegraphics[width=0.4\textwidth]{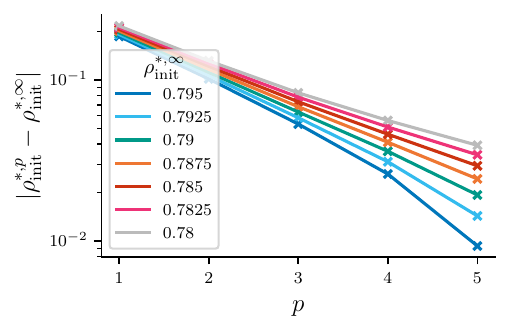}\includegraphics[width=0.4\linewidth]{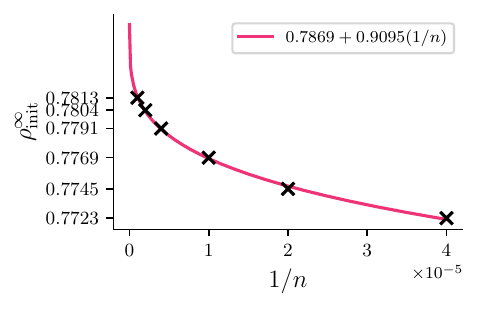}
   \caption{\textbf{Absolute majority rule $0011$ -- Extrapolation.} Extrapolation of the theoretical BDCM and empirical experiments to $n\to\infty$ and $p \to \infty$ respectively. \textit{(Left)} Scaling of the distance to different values of extrapolated $\rho^{*,\infty}_{\mathrm{init}}$ for data for $p=1,...,5$ for the BDCM. Assuming exponentially fast convergence in $p$, a $\rho^{*,\infty}_{\mathrm{init}}\sim0.7875\pm0.005$ is reasonable. \textit{(Right)} The location $\rhoinit$ of the slowest average convergence for experiments over $4,096$ samples of random regular graphs and initial configurations for a given graph size $n$ are recorded, and then extrapolated to $n\to\infty$. With this we estimate that in the large system limit, the transition happens at $\sim0.785\pm0.005$.}
   \label{app:fig:0011-supplements}
\end{figure*}

\begin{figure*}
   \centering
   \includegraphics[width=0.4\textwidth]{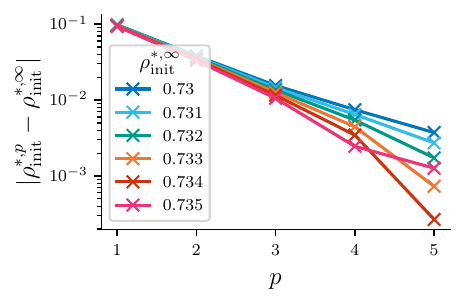}
   \caption{\textbf{Stubborn independent GCA $00$$+$$11$ -- Extrapolation.} As in Fig.\ref{app:fig:0011-supplements} we show different extrapolations of the BDCM's predictions for dynamical phase transitions for fixed $p$ and extrapolate them to $p\to\infty$, concluding that convergence within $\sim0.73\pm0.005$.
   The extrapolation for the empirics is taken from \cite{behrens2023backtracking}.}
   \label{app:fig:00+11-supplements}
\end{figure*}

\end{document}